\documentclass[aps,prd,twocolumn,superscriptaddress,showpacs,preprintnumbers,amsmath,amssymb,floatfix]{revtex4-1}

\usepackage[dvipdfmx]{graphicx}
\usepackage{dcolumn}
\usepackage{color}
\usepackage[mathlines]{lineno}
\usepackage{tabularx}
\usepackage{tikz}   
\usepackage{tabu}
\usepackage{tikz-feynman}
\usepackage{gensymb}
\usepackage{mathtools}
\usepackage[graphicx]{realboxes}
\usepackage{adjustbox}
\graphicspath{{ps}}
\usepackage{siunitx}
\renewcommand{\arraystretch}{1.1}
\usepackage{rotating} 
\newcolumntype{Y}{>{\centering\arraybackslash}X}
\begin{document}


\title{Measurement of the CKM Matrix Element $|V_{cb}|$  from  $B^{0} \to D^{*-} \ell^+ \nu_\ell$ at Belle}
\noaffiliation
\affiliation{University of the Basque Country UPV/EHU, 48080 Bilbao}
\affiliation{Beihang University, Beijing 100191}
\affiliation{University of Bonn, 53115 Bonn}
\affiliation{Brookhaven National Laboratory, Upton, New York 11973}
\affiliation{Budker Institute of Nuclear Physics SB RAS, Novosibirsk 630090}
\affiliation{Faculty of Mathematics and Physics, Charles University, 121 16 Prague}
\affiliation{Chonnam National University, Kwangju 660-701}
\affiliation{University of Cincinnati, Cincinnati, Ohio 45221}
\affiliation{Deutsches Elektronen--Synchrotron, 22607 Hamburg}
\affiliation{Duke University, Durham, North Carolina 27708}
\affiliation{Key Laboratory of Nuclear Physics and Ion-beam Application (MOE) and Institute of Modern Physics, Fudan University, Shanghai 200443}
\affiliation{Justus-Liebig-Universit\"at Gie\ss{}en, 35392 Gie\ss{}en}
\affiliation{II. Physikalisches Institut, Georg-August-Universit\"at G\"ottingen, 37073 G\"ottingen}
\affiliation{SOKENDAI (The Graduate University for Advanced Studies), Hayama 240-0193}
\affiliation{Gyeongsang National University, Chinju 660-701}
\affiliation{Hanyang University, Seoul 133-791}
\affiliation{University of Hawaii, Honolulu, Hawaii 96822}
\affiliation{High Energy Accelerator Research Organization (KEK), Tsukuba 305-0801}
\affiliation{J-PARC Branch, KEK Theory Center, High Energy Accelerator Research Organization (KEK), Tsukuba 305-0801}
\affiliation{Forschungszentrum J\"{u}lich, 52425 J\"{u}lich}
\affiliation{IKERBASQUE, Basque Foundation for Science, 48013 Bilbao}
\affiliation{Indian Institute of Technology Guwahati, Assam 781039}
\affiliation{Indian Institute of Technology Hyderabad, Telangana 502285}
\affiliation{Indian Institute of Technology Madras, Chennai 600036}
\affiliation{Indiana University, Bloomington, Indiana 47408}
\affiliation{Institute of High Energy Physics, Chinese Academy of Sciences, Beijing 100049}
\affiliation{Institute of High Energy Physics, Vienna 1050}
\affiliation{Institute for High Energy Physics, Protvino 142281}
\affiliation{INFN - Sezione di Napoli, 80126 Napoli}
\affiliation{INFN - Sezione di Torino, 10125 Torino}
\affiliation{Advanced Science Research Center, Japan Atomic Energy Agency, Naka 319-1195}
\affiliation{J. Stefan Institute, 1000 Ljubljana}
\affiliation{Institut f\"ur Experimentelle Teilchenphysik, Karlsruher Institut f\"ur Technologie, 76131 Karlsruhe}
\affiliation{Kennesaw State University, Kennesaw, Georgia 30144}
\affiliation{King Abdulaziz City for Science and Technology, Riyadh 11442}
\affiliation{Department of Physics, Faculty of Science, King Abdulaziz University, Jeddah 21589}
\affiliation{Kitasato University, Sagamihara 252-0373}
\affiliation{Korea Institute of Science and Technology Information, Daejeon 305-806}
\affiliation{Korea University, Seoul 136-713}
\affiliation{Kyoto University, Kyoto 606-8502}
\affiliation{Kyungpook National University, Daegu 702-701}
\affiliation{LAL, Univ. Paris-Sud, CNRS/IN2P3, Universit\'{e} Paris-Saclay, Orsay}
\affiliation{\'Ecole Polytechnique F\'ed\'erale de Lausanne (EPFL), Lausanne 1015}
\affiliation{P.N. Lebedev Physical Institute of the Russian Academy of Sciences, Moscow 119991}
\affiliation{Liaoning Normal University, Dalian 116029}
\affiliation{Faculty of Mathematics and Physics, University of Ljubljana, 1000 Ljubljana}
\affiliation{Ludwig Maximilians University, 80539 Munich}
\affiliation{University of Malaya, 50603 Kuala Lumpur}
\affiliation{University of Maribor, 2000 Maribor}
\affiliation{Max-Planck-Institut f\"ur Physik, 80805 M\"unchen}
\affiliation{School of Physics, University of Melbourne, Victoria 3010}
\affiliation{University of Mississippi, University, Mississippi 38677}
\affiliation{University of Miyazaki, Miyazaki 889-2192}
\affiliation{Moscow Physical Engineering Institute, Moscow 115409}
\affiliation{Moscow Institute of Physics and Technology, Moscow Region 141700}
\affiliation{Graduate School of Science, Nagoya University, Nagoya 464-8602}
\affiliation{Kobayashi-Maskawa Institute, Nagoya University, Nagoya 464-8602}
\affiliation{Universit\`{a} di Napoli Federico II, 80055 Napoli}
\affiliation{Nara Women's University, Nara 630-8506}
\affiliation{National Central University, Chung-li 32054}
\affiliation{National United University, Miao Li 36003}
\affiliation{Department of Physics, National Taiwan University, Taipei 10617}
\affiliation{H. Niewodniczanski Institute of Nuclear Physics, Krakow 31-342}
\affiliation{Nippon Dental University, Niigata 951-8580}
\affiliation{Niigata University, Niigata 950-2181}
\affiliation{University of Nova Gorica, 5000 Nova Gorica}
\affiliation{Novosibirsk State University, Novosibirsk 630090}
\affiliation{Osaka City University, Osaka 558-8585}
\affiliation{Pacific Northwest National Laboratory, Richland, Washington 99352}
\affiliation{Panjab University, Chandigarh 160014}
\affiliation{Peking University, Beijing 100871}
\affiliation{University of Pittsburgh, Pittsburgh, Pennsylvania 15260}
\affiliation{Theoretical Research Division, Nishina Center, RIKEN, Saitama 351-0198}
\affiliation{University of Science and Technology of China, Hefei 230026}
\affiliation{Seoul National University, Seoul 151-742}
\affiliation{Showa Pharmaceutical University, Tokyo 194-8543}
\affiliation{Soongsil University, Seoul 156-743}
\affiliation{University of South Carolina, Columbia, South Carolina 29208}
\affiliation{Sungkyunkwan University, Suwon 440-746}
\affiliation{School of Physics, University of Sydney, New South Wales 2006}
\affiliation{Department of Physics, Faculty of Science, University of Tabuk, Tabuk 71451}
\affiliation{Tata Institute of Fundamental Research, Mumbai 400005}
\affiliation{Department of Physics, Technische Universit\"at M\"unchen, 85748 Garching}
\affiliation{Toho University, Funabashi 274-8510}
\affiliation{Department of Physics, Tohoku University, Sendai 980-8578}
\affiliation{Earthquake Research Institute, University of Tokyo, Tokyo 113-0032}
\affiliation{Department of Physics, University of Tokyo, Tokyo 113-0033}
\affiliation{Tokyo Institute of Technology, Tokyo 152-8550}
\affiliation{Tokyo Metropolitan University, Tokyo 192-0397}
\affiliation{Virginia Polytechnic Institute and State University, Blacksburg, Virginia 24061}
\affiliation{Wayne State University, Detroit, Michigan 48202}
\affiliation{Yamagata University, Yamagata 990-8560}
\affiliation{Yonsei University, Seoul 120-749}
\author{E.~Waheed}\affiliation{School of Physics, University of Melbourne, Victoria 3010} 
\author{P.~Urquijo}\affiliation{School of Physics, University of Melbourne, Victoria 3010} 
  \author{I.~Adachi}\affiliation{High Energy Accelerator Research Organization (KEK), Tsukuba 305-0801}\affiliation{SOKENDAI (The Graduate University for Advanced Studies), Hayama 240-0193} 
  \author{K.~Adamczyk}\affiliation{H. Niewodniczanski Institute of Nuclear Physics, Krakow 31-342} 
  \author{H.~Aihara}\affiliation{Department of Physics, University of Tokyo, Tokyo 113-0033} 
  \author{S.~Al~Said}\affiliation{Department of Physics, Faculty of Science, University of Tabuk, Tabuk 71451}\affiliation{Department of Physics, Faculty of Science, King Abdulaziz University, Jeddah 21589} 
  \author{D.~M.~Asner}\affiliation{Brookhaven National Laboratory, Upton, New York 11973} 
  \author{H.~Atmacan}\affiliation{University of South Carolina, Columbia, South Carolina 29208} 
  \author{T.~Aushev}\affiliation{Moscow Institute of Physics and Technology, Moscow Region 141700} 
  \author{R.~Ayad}\affiliation{Department of Physics, Faculty of Science, University of Tabuk, Tabuk 71451} 
% ...
\collaboration{The Belle Collaboration}
 
\begin{abstract}
We present a new measurement of the CKM matrix element $|V_{cb}|$ from $B^{0} \to D^{*-} \ell^+ \nu_\ell$ decays, reconstructed with the full Belle data set of $711 \, \rm fb^{-1}$ integrated luminosity. Two form factor parameterizations, originally conceived by the Caprini-Lellouch-Neubert (CLN) and the Boyd, Grinstein and Lebed (BGL) groups, are used to extract the  product $\mathcal{F}(1)\eta_{\rm EW}|V_{cb}|$ and the decay form factors, where $\mathcal{F}(1)$ is the normalization factor and $\eta_{\rm EW}$ is a small electroweak correction. In the CLN parameterization we find $\mathcal{F}(1)\eta_{\rm EW}|V_{cb}| = (35.06 \pm 0.15 \pm 0.56) \times 10^{-3}$, $\rho^{2}=1.106 \pm 0.031 \pm 0.007$, $R_{1}(1)=1.229 \pm 0.028 \pm 0.009$, $R_{2}(1)=0.852 \pm 0.021 \pm 0.006$. For the BGL parameterization we obtain $\mathcal{F}(1)\eta_{\rm EW}|V_{cb}|= (34.93 \pm 0.23 \pm 0.59)\times 10^{-3}$, which is consistent with the World Average when correcting for $\mathcal{F}(1)\eta_{\rm EW}$. The branching fraction of $B^{0} \to D^{*-} \ell^+ \nu_\ell$ is measured to be  $\mathcal{B}(B^{0}\rightarrow D^{*-}\ell^{+}\nu_{\ell}) = (4.90 \pm 0.02  \pm 0.16)\%$. We also present a new test of lepton flavor universality violation in semileptonic $B$ decays, $\frac{{\cal B }(B^0 \to D^{*-} e^+ \nu)}{{\cal B }(B^0 \to D^{*-} \mu^+ \nu)} = 1.01 \pm 0.01 \pm 0.03~$.
The errors correspond to the statistical and systematic uncertainties respectively. This is the most precise measurement of $\mathcal{F}(1)\eta_{\rm EW}|V_{cb}|$ and form factors to date and the first experimental study of the BGL form factor parameterization in an experimental measurement.
\end{abstract}

\maketitle
\section{Introduction}
    The decay $B^{0} \to D^{*-} \ell^+ \nu_\ell$ is used to measure the Cabibbo-Kobayashi-Maskawa (CKM) matrix element $|V_{cb}|$ ~\cite{cite-ckm1,cite-ckm2}, the magnitude of the coupling between $b$ and $c$ quarks in  weak interactions, and is a fundamental parameter of the Standard Model (SM). The $B^{0} \to D^{*-} \ell^+ \nu_\ell$ decay is studied in the context of Heavy Quark Effective Theory (HQET) in which the hadronic matrix elements are parameterized by the form factors that can describe this decay. The decay amplitudes of $B^{0} \to D^{*-} \ell^+ \nu_\ell$ are described by three helicity amplitudes which are extracted from the three polarization states of the $D^{*}$ meson: two transverse polarisation terms, $H_{\pm}$, and one longitudinal polarisation term, $H_{0}$. 

There exists a long standing tension in the measurement of $|V_{cb}|$ using the inclusive approach, based on the decay mode $B \to X_{c}\ell \nu$ and the exclusive approach with $B \to D^* \ell \nu$. Currently, the world averages for $|V_{cb}|$ for inclusive and exclusive decay modes are~\cite{cite-hflav}:
\begin{eqnarray}
|V_{cb}| &=& (42.2 \pm 0.8)  \times 10^{-3} ~(\text{inclusive}),\label{eq:inclusive_val}  \\
|V_{cb}| &=& (39.1 \pm 0.4)  \times 10^{-3} ~(\text{CLN, exclusive}),\label{eq:cln_exclusive_val}
\end{eqnarray}
where the errors are the experimental and the theoretical combined. The difference between the inclusive and exclusive approaches is more than $2.5\sigma$. It is thought that the previous theoretical approaches using the CLN form factor parameterization~\cite{cite-CLN} were model dependent and introduced a bias, and therefore model independent form factor approaches based on BGL~\cite{cite-BGL} should be used. In this paper we report data fits with both approaches for the first time.
In this paper, the decay is reconstructed in the channel:  $B^{0} \to D^{*-} \ell^+ \nu_\ell$ followed by $D^{*-}\rightarrow {\bar{D}}^{0}\pi_{s}^{-}$ and   $\bar{D^{0}}\rightarrow K^{-}\pi^{+}$~\cite{cite-chargeconjugation}. This channel offers the best purity for the measurement, which is critical as the measurement will be limited by systematic uncertainties. This is the most precise determination of $|V_{cb}|$ performed with exclusive semileptonic $B$ decays to date.  
This result supersedes the previous results on  $B^{0} \to D^{*-} \ell^+ \nu_\ell$ with an untagged approach from Belle ~\cite{cite-dungel}. A major experimental improvement to the efficiency of the track reconstruction software was implemented in 2011, leading to substantially higher slow pion tracking efficiencies~\cite{tracking-efficiency} and hence much larger signal yields than in the previous analysis.

\section{Experimental Apparatus and data samples}\label{data-section}
We use the full $\Upsilon(4S)$ data sample containing $( 772\pm 11 )  \times 10^6  B\bar{B}$ pairs equivalent to $711 \, \rm fb^{-1}$ of integrated luminosity recorded with the Belle detector~\cite{cite-Belle} at the asymmetric-energy $e^+ e^-$ collider KEKB~\cite{cite-KEKB}.  An additional 88 fb$^{-1}$ of data  collected 60 MeV below the $\Upsilon(4S)$ was used for the estimation of $q{\bar q}$ ($q = u, d, s, c$) continuum background.

The Belle detector is a large-solid-angle magnetic spectrometer that consists of a silicon vertex detector (SVD), a 50-layer central drift chamber (CDC), an array of aerogel threshold Cherenkov counters (ACC), a barrel-like arrangement of time-of-flight scintillation counters (TOF), and an electromagnetic calorimeter (ECL) comprised of CsI(Tl) crystals located inside a superconducting solenoid coil that provides a 1.5~T magnetic field. An iron flux-return located outside of the coil is instrumented to detect $K_L^0$ mesons and to identify muons (KLM). The detector is described in detail elsewhere~\cite{cite-Belle}. Two inner detector configurations were used. A 2.0~cm radius beampipe and a 3-layer silicon vertex detector was used for the first subsample of $152 \times 10^6 B\bar{B}$ pairs (denoted as SVD1), while a 1.5~cm radius beampipe, a 4-layer silicon detector and a small-cell inner drift chamber were used to record the remaining $620 \times 10^6 B\bar{B}$ pairs~\cite{cite-SVD2} (denoted as SVD2). We refer to these subsamples later in the paper.

\subsection{Monte Carlo Simulation}

Monte Carlo simulated events are used to determine the analysis selection criteria, study the background and estimate the signal reconstruction efficiency. Events with a $B \bar{B}$ pair are generated using EvtGen~\cite{cite-EvtGen}, and the $B$ meson decays are reproduced based on branching fractions reported in Ref.~\cite{cite-PDG2016}. The hadronization process of $B$ meson decays that do not have  experimentally-measured branching fractions is inclusively reproduced by PYTHIA~\cite{cite-PYTHIA}. For continuum events, the initial quark pair is hadronized by PYTHIA, and hadron decays are modelled by EvtGen. The final-state radiation from charged particles is added using PHOTOS~\cite{cite-PHOTOS}. Detector responses are simulated with GEANT3~\cite{cite-GEANT}. 

\subsection{Event Reconstruction and Selection Criteria}

Charged particle tracks are required to originate from the interaction point, and to have good track fit quality. The criteria for the track impact parameters in the $r-\phi$ and $z$ directions are: $dr<2$ cm and $|dz|<4$ cm, respectively. In addition we require that each track has at least one associated hit in any layer of the SVD. For pion and kaon candidates, we use likelihoods determined using the Cherenkov light yield in the ACC, the time-of-flight information from the TOF, and $dE/dx$ from the CDC.

Neutral $\bar{D}^0$ meson candidates are reconstructed in the clean $\bar{D}^0\to K^- \pi^+$ decay channel. The daughter tracks are fitted to a common vertex using a Kalman fit algorithm, with a $\chi^2$-probability requirement of greater than $10^{-3}$ to reject misreconstructed  $\bar{D}^0$ candidates. The reconstructed $\bar{D}^0$ mass is required to be in a window of $\pm13.75$ MeV/$c^2$ from the nominal $D^0$ mass, corresponding to a width of 2.5 $\sigma$, determined from data.

The $\bar{D}^0$ candidates are combined with an additional pion that has a charge opposite that of the kaon, to form $D^{*-}$ candidates. Pions produced in this transition are close to the kinematic threshold, with a mean momentum of approximately 100 MeV/$c$, hence are denoted slow pions, $\pi_s^-$. There are no SVD hit requirements for slow pions. Another vertex fit is performed between the $D^0$ and the $\pi_s^-$ and a $\chi^2$-probability requirement of greater than $10^{-3}$ is again imposed.  The invariant mass difference between the $D^{*-}$ and the $\bar{D}^{0}$ candidates, $\Delta M = M_{D^{*}}-M_{D^{0}}$,  is first required to be less than 165 MeV/$c^2$ for the background fit, and further tightened for the signal yield determination.  

\begin{figure}[htb]
 \centering
 \includegraphics[scale=0.3]{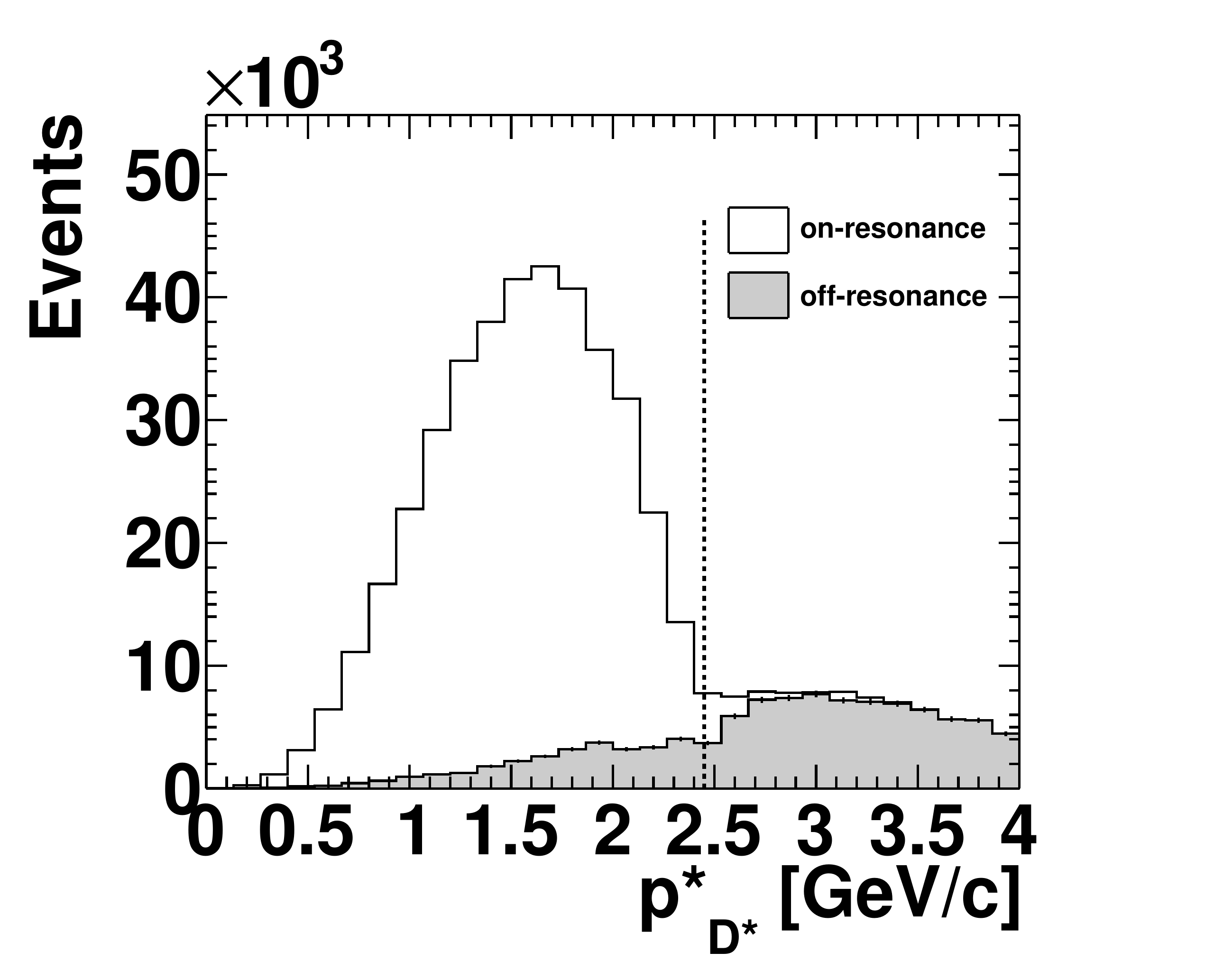} 
 \caption{The $D^*$ momenta in the CM frame, for on-resonance and scaled off-resonance data. The dotted line shows the cut applied for suppression of continuum.}
 \label{fig:pdstar}
 \end{figure}

Although the contribution from continuum is relatively small in this analysis and it is dominated by charm by fake $D^*$, we further suppress prompt charm by imposing an upper threshold on the $D^*$ momentum of 2.45 GeV/$c$ in the center-of-mass (CM) frame (Fig.~\ref{fig:pdstar}).

Candidate $B$ mesons are reconstructed by combining $D^{*}$ candidates with an oppositely charged electron or muon. Electron candidates are identified using the ratio of the energy detected in the ECL to the momentum of the track, the ECL shower shape, the distance between the track at the ECL surface and the ECL cluster centre, the energy loss in the CDC ($dE/dx$) and the response of the ACC. For electron candidates we search for nearby bremsstrahlung photons in a cone of 3 degrees around the electron track, and sum the momenta with that of the electron. Muons are identified by their penetration range and transverse scattering in the KLM system. In the momentum region relevant to this analysis, charged leptons are identified with an efficiency of about 90\%, while the probabilities to misidentify a pion as an electron and muon are 0.25\% and 1.5\% respectively~\cite{cite-emuid1} \cite{cite-emuid2}.
We impose lower thresholds on the momentum of the leptons, such that they reach the respective particle identification detectors for good hadron fake rejection. Here we impose lab frame momentum thresholds of 0.3 GeV/$c$ for electrons and 0.6 GeV/$c$ for muons. We furthermore require an upper threshold of 2.4 GeV/$c$ in the CM frame  to reject continuum events.

\section{Decay Kinematics}
\begin{figure}[htb] 
 \centering
\begin{tikzpicture} \begin{feynman}
\vertex (a1) {\(\overline b\)}; \vertex[right=1.5cm of a1] (a2);
\vertex[right=1cm of a2] (a3) ;
\vertex[right=1.5cm of a3] (a4) {\(\overline c\)};
\vertex[below=2em of a1] (b1) {\(d\)}; 
\vertex[below=2em of a4] (b2) {\(d\)};
\vertex at ($(a2)!0.5!(a3)!0.0cm!90:(a3)$) (d); 
\vertex[above=2em of a4] (c1) {\(\nu_{\ell}\)}; 
\vertex[above=2em of c1] (c3) {\(\ell^+\)}; 
\vertex at ($(c1)!0.5!(c3) - (1cm, 0)$) (c2);
\diagram* {
(a4) -- [fermion] (a3) -- [fermion] (a1),
(b1) -- [fermion] (b2),
(c3) -- [fermion, out=180, in=45] (c2) -- [fermion, out=-45, in=180] (c1), 
(d) -- [blue, boson, edge label=\(W^{+}\)] (c2),
};
\draw [decoration={brace}, decorate] (b1.south west) -- (a1.north west) node [pos=0.5, left] {\(B^{0}\)};
\draw [decoration={brace}, decorate] (a4.north east) -- (b2.south east) node [pos=0.5, right] {\(D^{*-}\)};
\end{feynman} \end{tikzpicture} \caption{Tree level Feynman diagram for $B^{0} \rightarrow D^{*-}\ell^{+}\nu_\ell$.}
 \label{tree}
 \end{figure}
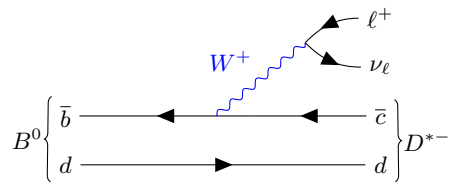
 
\begin{figure}[ht!]
 \centering
 \includegraphics[scale=0.2]{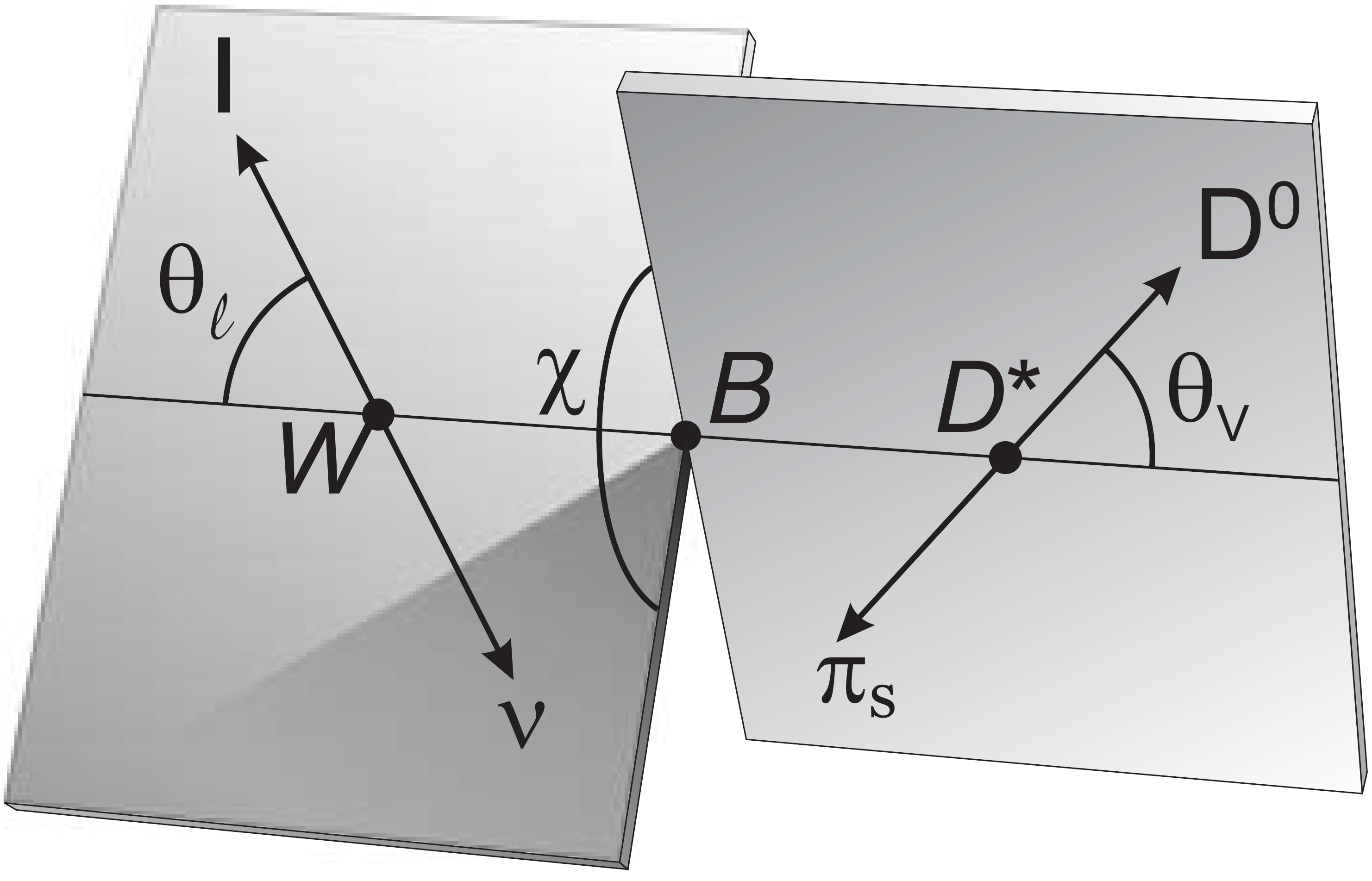} 
 \caption{Definition of the angles $\theta_{\ell}$, $\theta_{\rm v}$ and $\chi$ for the decay $B^{0}\rightarrow D^{*-}\ell^{+}\nu_{\ell}$.}
 \label{angle}
 \end{figure}

The tree level transition of the  $B^{0}\rightarrow D^{*-}\ell^{+}\nu_{\ell}$ decay is shown in Fig.~\ref{tree}. Three angular variables and the hadronic recoil are used to describe this decay. The latter is defined as follows:  
\begin{equation}
w  = \frac{P_{B}\cdot P_{D^{*}}}{m_{B}m_{D^{*}}} = \frac{m^{2}_{B}+ m^{2}_{D^{*}}-q^{2}}{2 m_{B}m_{D^{*}}},
\end{equation}
where $P_{B}$ and $P_{D^{*}}$ are four momenta of  the $B$ and the $D^{*}$ mesons respectively,  $m_{B}$, $m_{D^{*}}$ are their masses, and $q^{2}$ is the invariant mass squared of the lepton-neutrino system. The range of $w$ is restricted by the allowed values of $q^{2}$ such that the minimum value of $q^2_{\rm min} = m_\ell^2 \approx 0~ \rm GeV^2$ corresponds to the maximum value of $w$,
\begin{equation}
w_{\rm max} = \frac {m^{2}_{B}+m^{2}_{D^{*}}} {2 m_{B}m_{D^{*}}}.
\end{equation}
The three angular variables are depicted in Fig.~\ref{angle} and are defined as follows:
 \begin{itemize}
  \item \textbf{$\theta_{\ell}$}: the angle between the direction of the lepton and the direction opposite the $B$ meson in the virtual $W$ rest frame.
  \item \textbf{$\theta_{\rm v}$}: the angle between the direction of the $D^{0}$ meson and the direction opposite the $B$ meson in the $D^{*}$ rest frame. 
\item \textbf{$\chi$}: the angle between the two planes formed by the decays of the $W$ and the $D^{*}$ meson, defined in the rest frame of the $B^{0}$ meson.
\end{itemize}

\section{Semileptonic decays}
In the massless lepton limit, the $B^{0} \to D^{*-} \ell^+ \nu_\ell$ differential decay rate is given by~\cite{cite-CLN}
\begin{eqnarray}\label{eq:diff}
&&\frac{d\Gamma (B^{0} \to D^{*-} \ell^+ \nu_\ell)}{dwd\cos\theta_\ell d\cos\theta_{\rm v}d\chi} = \nonumber\\
&& \frac{\eta_{\rm EW}^2 3 m_B m_{D^*}^2}{4(4\pi)^2} G_F^2 |V_{cb}|^2\sqrt{w^2-1}(1 - 2wr + r^2)   \nonumber \\
&&\left\{ ( 1 - \cos\theta_\ell)^2\sin^2\theta_{\rm v}H_+^2(w) + (1+\cos\theta_\ell)^2\sin^2\theta_{\rm v}H_-^2(w)  \right. \nonumber \\
&& +4\sin^2\theta_\ell\cos^2\theta_{\rm v}H_0^2 - 2\sin^2\theta_\ell\sin^2\theta_{\rm v}\cos2\chi H_+(w)H_-(w)  \nonumber \\
&& -4\sin\theta_\ell(1-\cos\theta_\ell)\sin\theta_{\rm v}\cos\theta_{\rm v}\cos\chi H_+(w)H_0(w) \nonumber \\
&& \left. +4\sin\theta_\ell(1+\cos\theta_\ell)\sin\theta_{\rm v}\cos\theta_{\rm v}\cos\chi H_-(w)-H_0(w) \right\},
\end{eqnarray}
where $r=m_{D^*}/m_B$, $G_{F} = (1.6637 \pm 0.00001) \times 10^{-5} \rm \hbar c^{2}\rm GeV^{-2}$ and $\eta_{\rm EW}$ is a small electroweak correction (Calculated to be 1.006 in Ref.~\cite{cite-etaew}).

\subsection{The CLN Parameterization}
The helicity amplitudes $H_{\pm,0}(w)$ in Eq.~\ref{eq:diff} are given in terms of three form factors. In the CLN parameterization~\cite{cite-CLN} one writes these helicity amplitudes in terms of the form factor $h_{A_1}(w)$ and the form factor ratios $R_{1,2}(w)$. They are defined as
\begin{eqnarray}\label{eq:cln}
h_{A_1}(w) &=& h_{A_1}(1)\left[ 1 - 8 \rho^2z + (53\rho^2 - 15)z^2  \right. \nonumber\\
&&\left. - (231\rho^2-91)z^3\right], \nonumber\\
R_1(w) &=& R_1(1) - 0.12 (w-1) + 0.05 (w-1)^2, \nonumber\\
R_2(w) &=& R_2(1) - 0.11 (w-1) - 0.06 (w-1)^2, 
\end{eqnarray}
where $z=(\sqrt{w+1}-\sqrt{2})/(\sqrt{w+1}+\sqrt{2})$. In addition to the form factor normalization, there are three independent parameters $\rho^2$, $R_{1}(1)$ and $R_{2}(1)$. The values of these parameters are not calculated theoretically instead they are extracted by an analysis of experimental data.

\subsection{The BGL Parameterization}
A more general parameterization comes from BGL~\cite{cite-BGL}, recently used in Refs.~\cite{,cite-grinstein,cite-gambino}. In their approach, the helicity amplitudes $H_i$ are given by
\begin{eqnarray}
H_0(w) &=& {\mathcal F}_1(w)/\sqrt{q^2}~,\nonumber\\
H_\pm(w) &=& f(w) \mp m_Bm_{D^*}\sqrt{w^2-1}g(w)~.
\end{eqnarray}

The three BGL form factors can be written as a series in powers of $z$,
\begin{eqnarray}\label{eq:bglff}
f(z) &=& \frac{1}{P_{1+}(z)\phi_f(z)}\sum_{n=0}^{\infty}a_n^fz^n~,\nonumber\\
{\mathcal F}_1(z) &=& \frac{1}{P_{1+}(z)\phi_{\mathcal F_1}(z)}\sum_{n=0}^{\infty}a_n^{{\mathcal F_1}}z^n~,\nonumber\\
g(z) &=& \frac{1}{P_{1-}(z)\phi_g(z)}\sum_{n=0}^{\infty}a_n^gz^n~.
\end{eqnarray}
In these equations the Blaschke factors, $P_{1\pm}$, are given by
\begin{eqnarray}
P_{1\pm}(z) = \prod_{P=1}^{n} \frac{z-z_P}{1-zz_P}~,
\end{eqnarray}
where $z_P$ is defined as
\begin{eqnarray}
z_P = \frac{\sqrt{t_+ - m_P^2}- \sqrt{t_+-t_-} }{\sqrt{t_+ - m_P^2} + \sqrt{t_+-t_-} }~,
\end{eqnarray}
while $t_{\pm}=(m_B\pm m_{D^*})^2$ and $m_{P}$ denotes the masses of the $B_c^*$ resonances. The product is extended to include all the $B_c$ resonances below the $B-D^*$ threshold of 7.29 $\rm GeV/c^2$ with the appropriate quantum numbers ($1^+$ for $f(w)$ and ${\cal F}_1(w)$, and $1^-$ for $g(w)$). We use the the $B_c$ resonances listed in Table~\ref{tab:bc}.
\begin{table}[htb]
\centering
\caption{The $B_c^{(*)}$ masses used in the Blaschke factors of the BGL parameterization.}
 \renewcommand{\arraystretch}{1.3}
\begin{tabularx}{0.5\linewidth}{lY}
\hline \hline
Type & Mass (GeV/$c^2$) \\
\hline
$1^-$ & 6.337\\
$1^-$ & 6.899\\
$1^-$ & 7.012\\
$1^-$ & 7.280\\
$1^+$ & 6.730\\
$1^+$ & 6.736\\
$1^+$ & 7.135\\
$1^+$ & 7.142\\
\hline
\end{tabularx}
\label{tab:bc}
\end{table}
The $B_c$ resonances also enter the $1^-$ unitarity bounds as single particle contributions. The outer functions $\phi_i$ for $i=g,~f,~{\cal F}_1$ are as follows:
\begin{eqnarray}
\phi_g(z) &=& \sqrt{\frac{n_I}{3 \pi \chi_{1-}^T(0)}} \nonumber \\
&& \times \frac{2^4r^2(1+z)^2(1-z)^{-1/2}}{\left[ (1+r)(1-z) + 2\sqrt{r}(1+z)\right]^4}~, \nonumber \\
\phi_f(z) &=& \frac{4r}{m_B^2}\sqrt{\frac{n_I}{3 \pi \chi_{1+}^T(0)}} \nonumber\\
&&\times \frac{(1+z)(1-z)^{3/2}}{\left[ (1+r)(1-z) + 2\sqrt{r}(1+z)\right]^4}~, \nonumber \\
\phi_{\mathcal F_1}(z) &=& \frac{4r}{m_B^3}\sqrt{\frac{n_I}{6 \pi \chi_{1+}^T(0)}} \nonumber\\
&&\times \frac{(1+z)(1-z)^{5/2}}{\left[ (1+r)(1-z) + 2\sqrt{r}(1+z)\right]^5}~,  
\end{eqnarray}
where $\chi_{1+}^T(0)$ and $\chi_{1-}^T(0)$ are constants given in Table~\ref{tab:chit}, and $n_I=2.6$ represents the number of spectator quarks (three), decreased by a large and conservative SU(3) breaking factor. 
\begin{table}[htb]
\centering
\caption{Inputs used in the BGL fit.}
 \renewcommand{\arraystretch}{1.3}
\begin{tabularx}{0.6\linewidth}{lr}
\hline \hline
Input & Value \\
\hline
$m_{B^0}$ 			& $5.279$ GeV$/c^2$\\
$m_{D^{*+}}$ 			& $2.010$ GeV$/c^2$\\
$\eta_{EW}$ 			& $1.0066$\\
$\chi^T_{1+}(0)$	~~& $5.28 \times 10^{-4}$ $\rm (GeV/c{^2})^{-2}$\\
$\chi^T_{1-}(0)$ 	~~& $3.07 \times 10^{-4}$ $\rm (GeV/c{^2})^{-2}$\\
\hline
\end{tabularx}
\label{tab:chit}
\end{table}
At zero recoil ($w=1$ or $z=0$) there is a relation between two of the form factors,
\begin{eqnarray}
{\mathcal F_1}(0) = (m_B - m_{D^*})f(0).
\end{eqnarray}
The coefficients of the expansions in Eq.~\ref{eq:bglff} are subject to unitarity bounds based on analyticity and the operator product expansion applied to correlators of two hadronic $\bar c b$ currents:
\begin{eqnarray}
\sum_{i=0}^{\infty} (a_n^g)^2 <1~,\nonumber\\
\sum_{i=0}^{\infty} \left[ (a_n^f)^2 + (a_n^{\mathcal F_1})^2 \right] <1~.
\end{eqnarray}
They ensure rapid convergence of the $z$ expansion over the whole physical region, $0<z<0.056$. The series must be truncated at some power $\rm n_{max}$. 

\section{Background estimation}
The most powerful discriminator against background is the cosine of the angle between the $B$ and the $D^*\ell$ momentum vectors in the CM frame under the assumption that the $B$ decays to $D^* \ell \nu$. In the CM frame, the $B$ direction lies on a cone around the $D^*\ell$ axis with an opening angle $2\cos\theta_{B,D^*\ell}$, defined as:
\begin{eqnarray}
\cos\theta_{B,D^*\ell} = \frac{2E_B^*E_{D^*\ell}^* - m_B^2 - m_{D^*\ell}^2}{2|\vec p_B^*||\vec p_{D^*\ell}^*|},
\end{eqnarray}
where $E_B^*$ is half of the CM energy and $|\vec p_{B}^*|$ is $\sqrt{E_B^{*2}-m_B^2}$. The quantities $E_{D^*\ell}^*$, $\vec p^*_{D^*\ell}$ and $m_{D^*\ell}$ are determined from the reconstructed $D^*\ell$ system.

The remaining background in the sample is split into the following categories.
\begin{itemize}
\item $B \to D^{**} \ell \nu$, both resonant where $D^{**}$ decays to a $D^*$, and nonresonant $B \to D^* \pi \ell \nu$ decays.
\item Correlated cascade decays where the $D^*$ and $\ell$ originate from the same $B$, e.g. $B \to D^{*} \tau \nu$ ($\tau \to \ell \nu \bar \nu$), and $B \to D^{*} D$, $D\to \ell X$.
\item Uncorrelated decays, where the $D^*$ and $\ell$ originate from different $B$ mesons in the event.
\item Mis-identified leptons (fake leptons): the probability for a hadron being identified as a lepton is small but not negligible in the low momentum region, and is higher for muons.
\item Fake $D^*$ candidates, where the $D^*$ is incorrectly reconstructed.
\item $q \bar q$ continuum, typically $e^+e^- \to c\bar c$.
\end{itemize}

To model the $B \to D^{**} \ell \nu$ component, which is comprised of four $P$-wave resonant modes ($D_1$, $D_0^*$, $D_1^\prime$, $D_2^*$) for both neutral and charged $B$ decays, we correct the branching fractions and form factors. The $P$-wave charm mesons are categorized according to the angular momentum of the light constituent, $j_\ell$, namely the $j_\ell^P=1/2^-$ doublet of $D_0^*$ and $D_1^\prime$ and the $j_\ell^P=3/2^-$ doublet $D_1$ and $D_2^*$. The shapes of the $B \to D^{**} \ell \nu$ $q^2$ distributions are corrected to match the predictions of the LLSW model~\cite{cite-LLSW}.
An additional contribution from nonresonant modes is considered, although the rate appears to be consistent with zero in recent measurements~\cite{cite-dstarpilnu}.

\begin{figure*}[htb]
 \centering
 \includegraphics[width=0.48\linewidth]{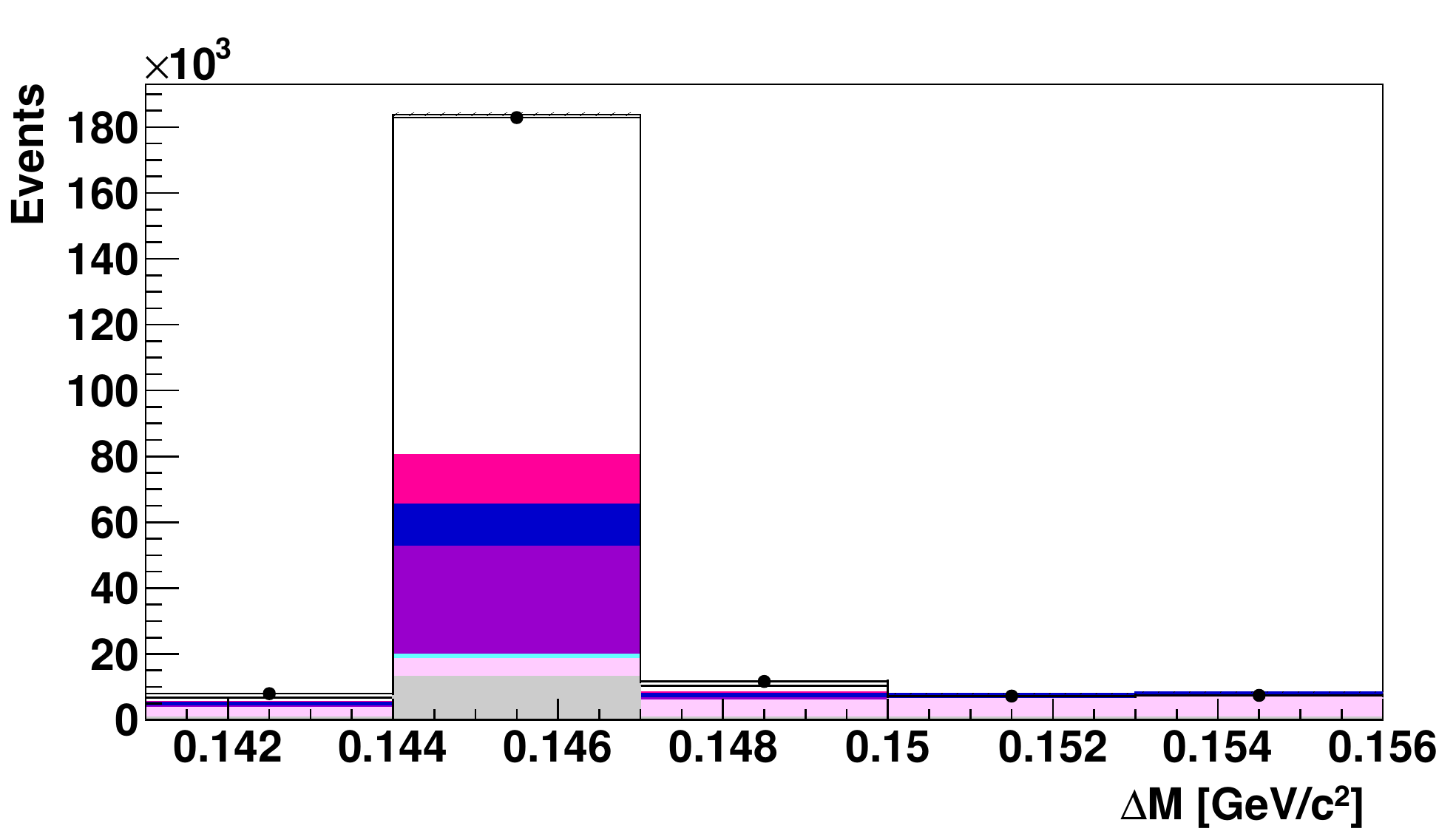} 
 \includegraphics[width=0.48\linewidth]{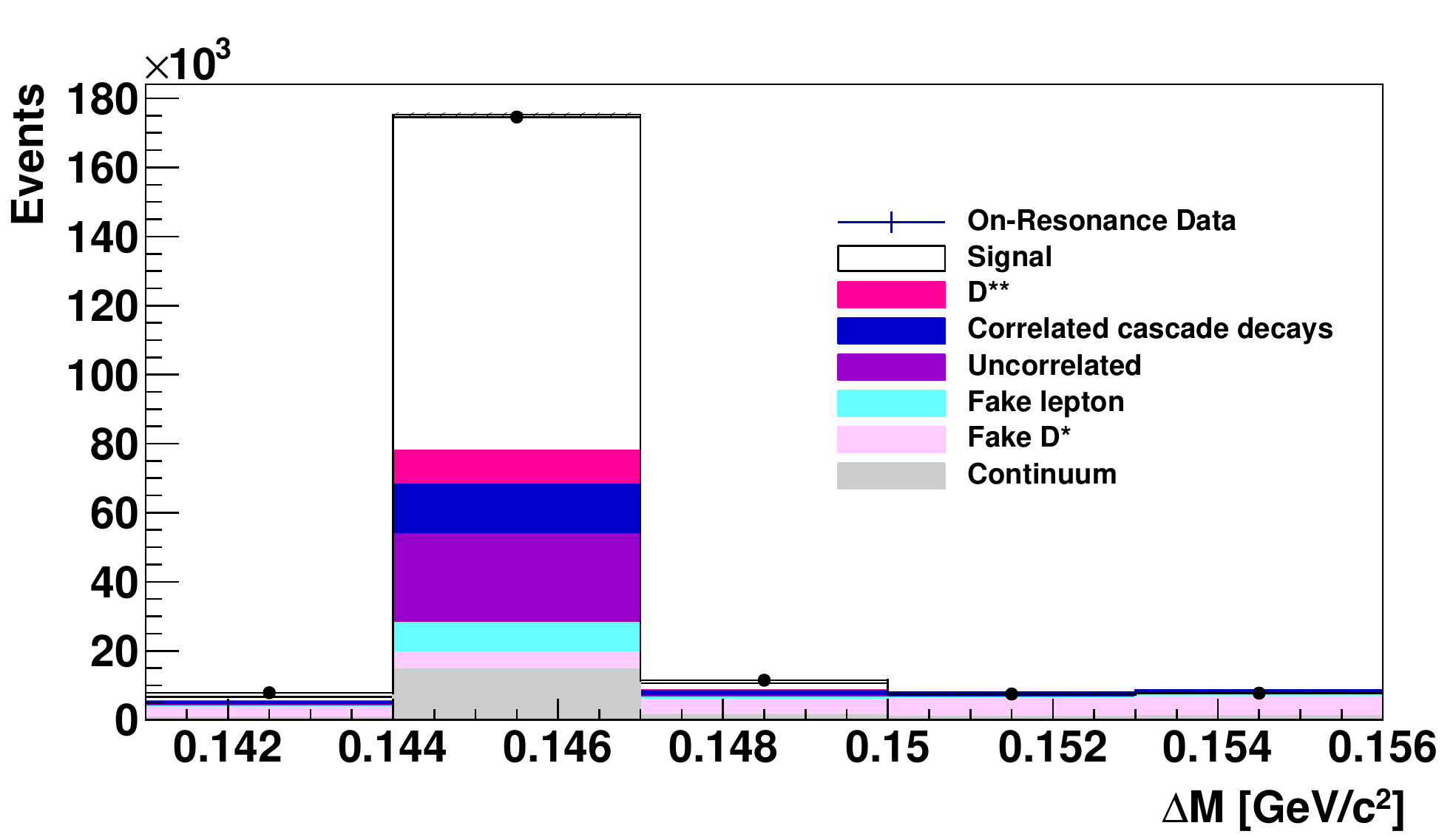} 
 \includegraphics[width=0.48\linewidth]{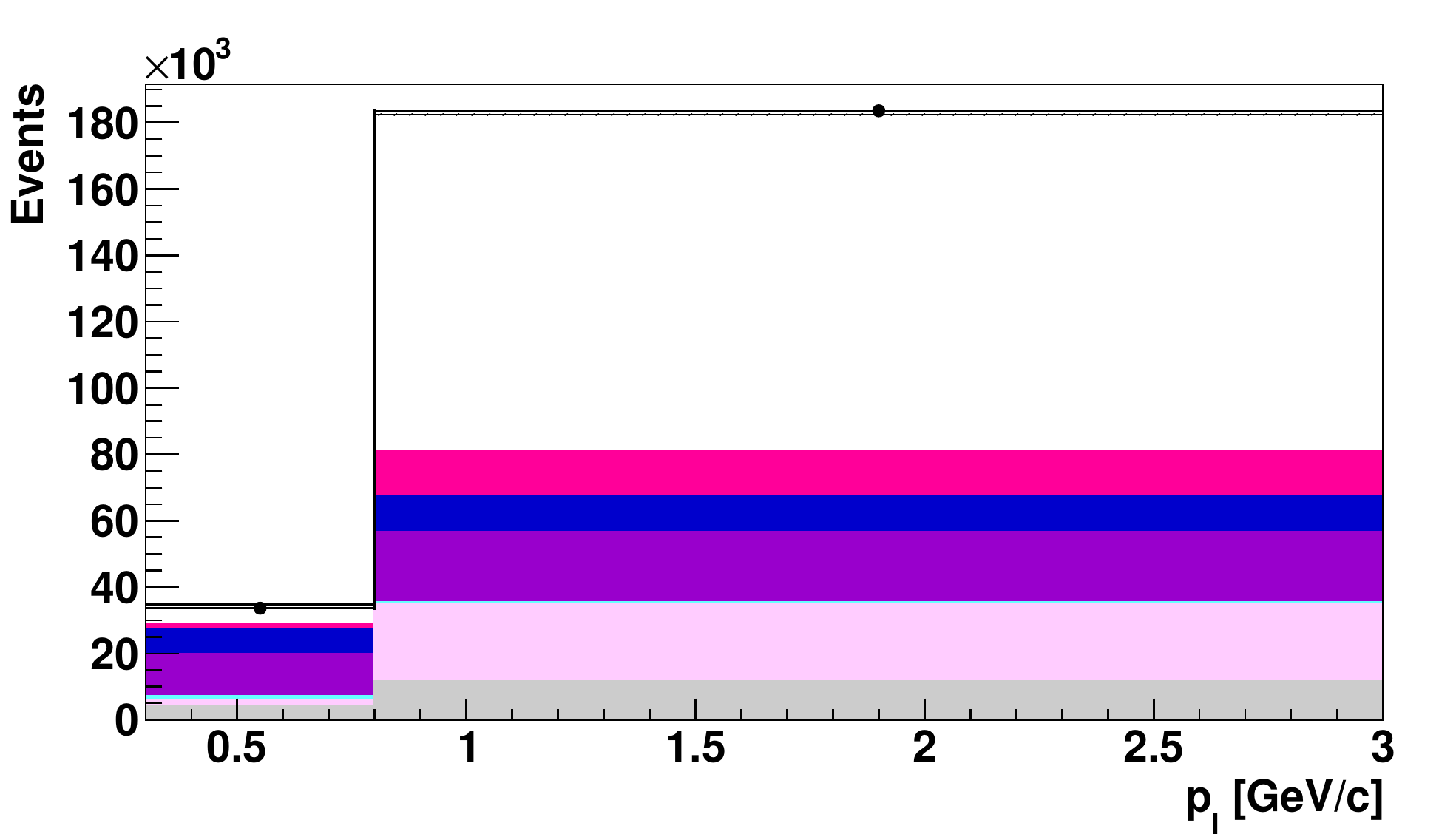} 
 \includegraphics[width=0.48\linewidth]{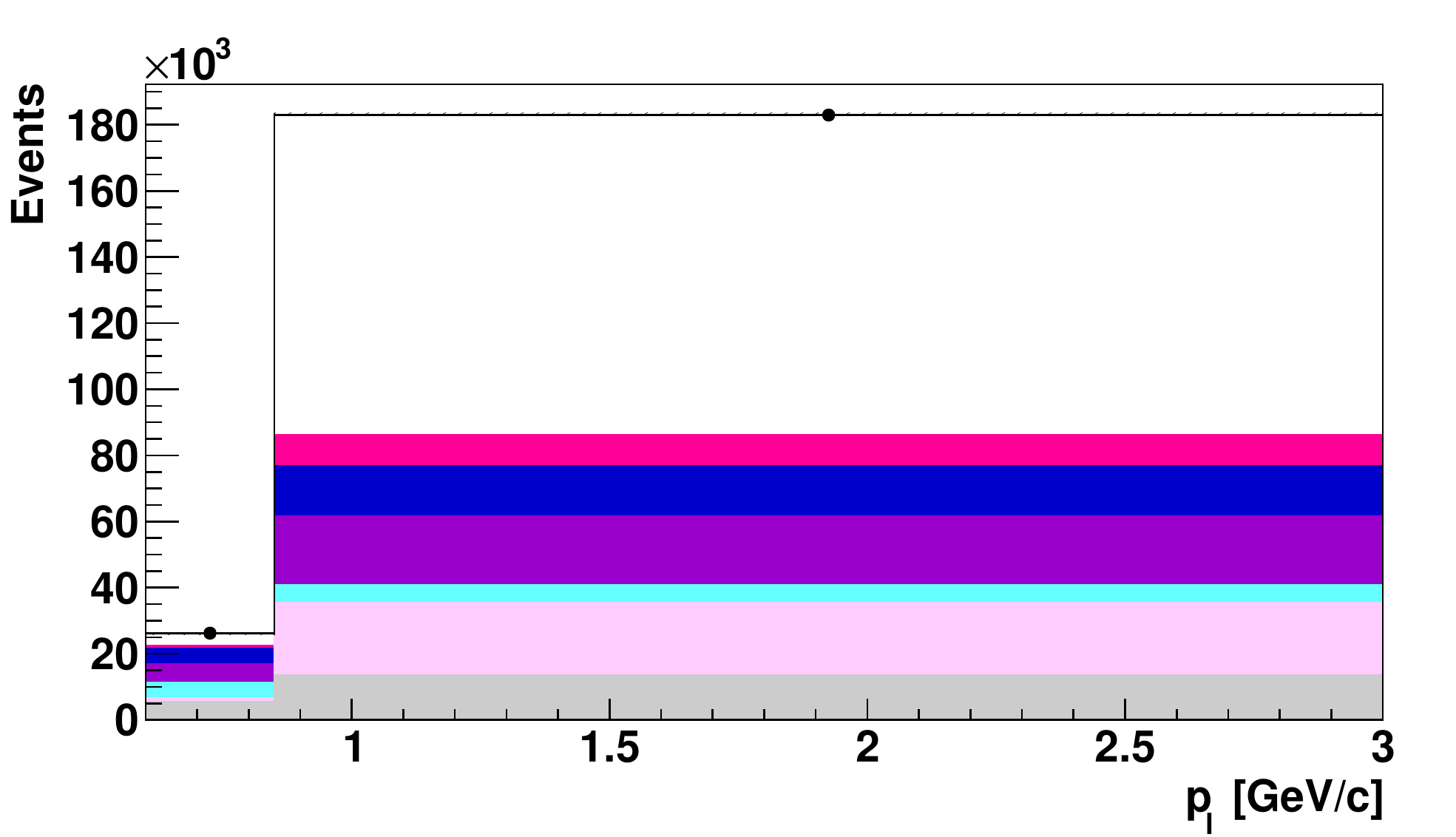} 
 \includegraphics[width=0.48\linewidth]{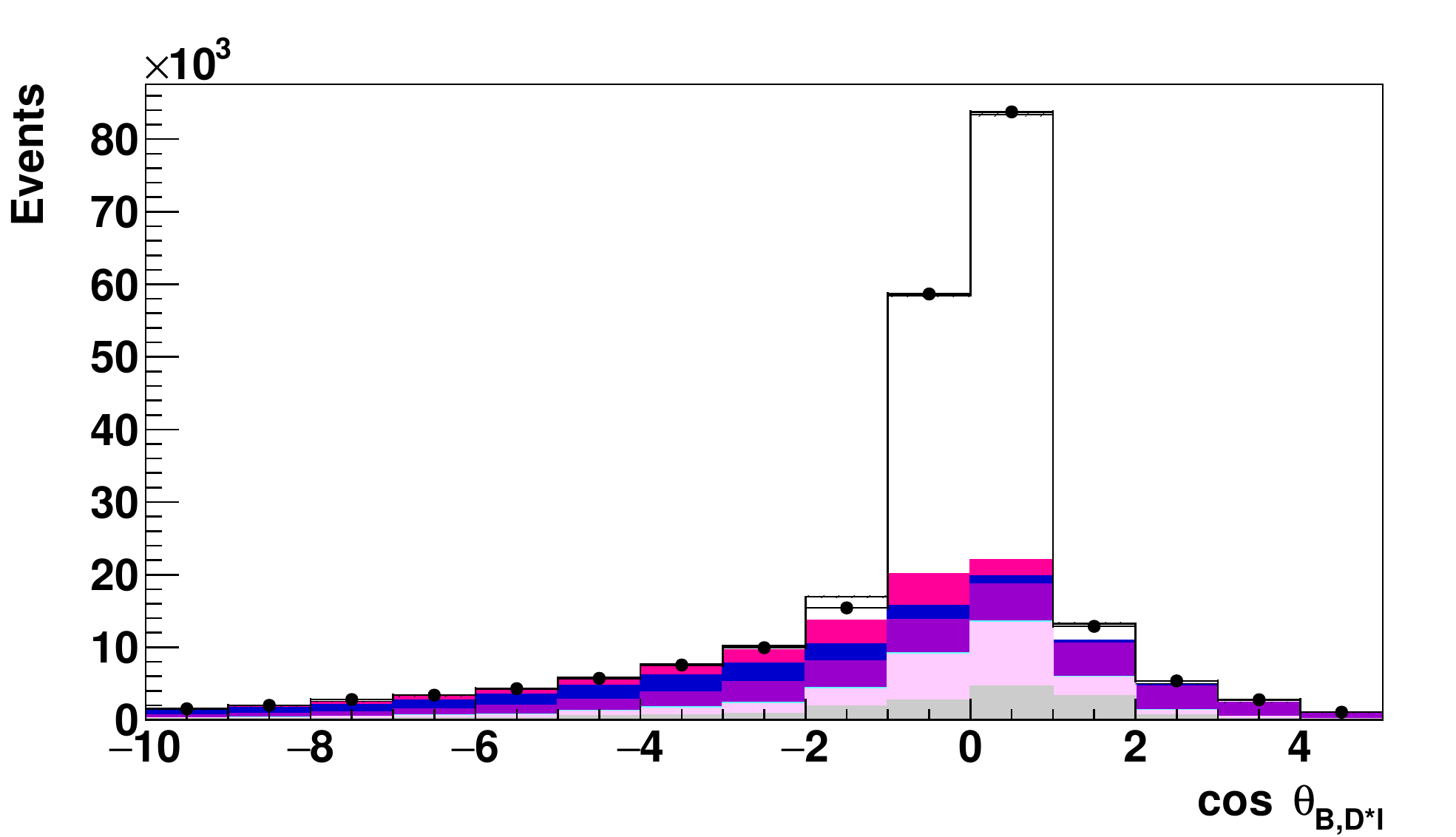} 
 \includegraphics[width=0.48\linewidth]{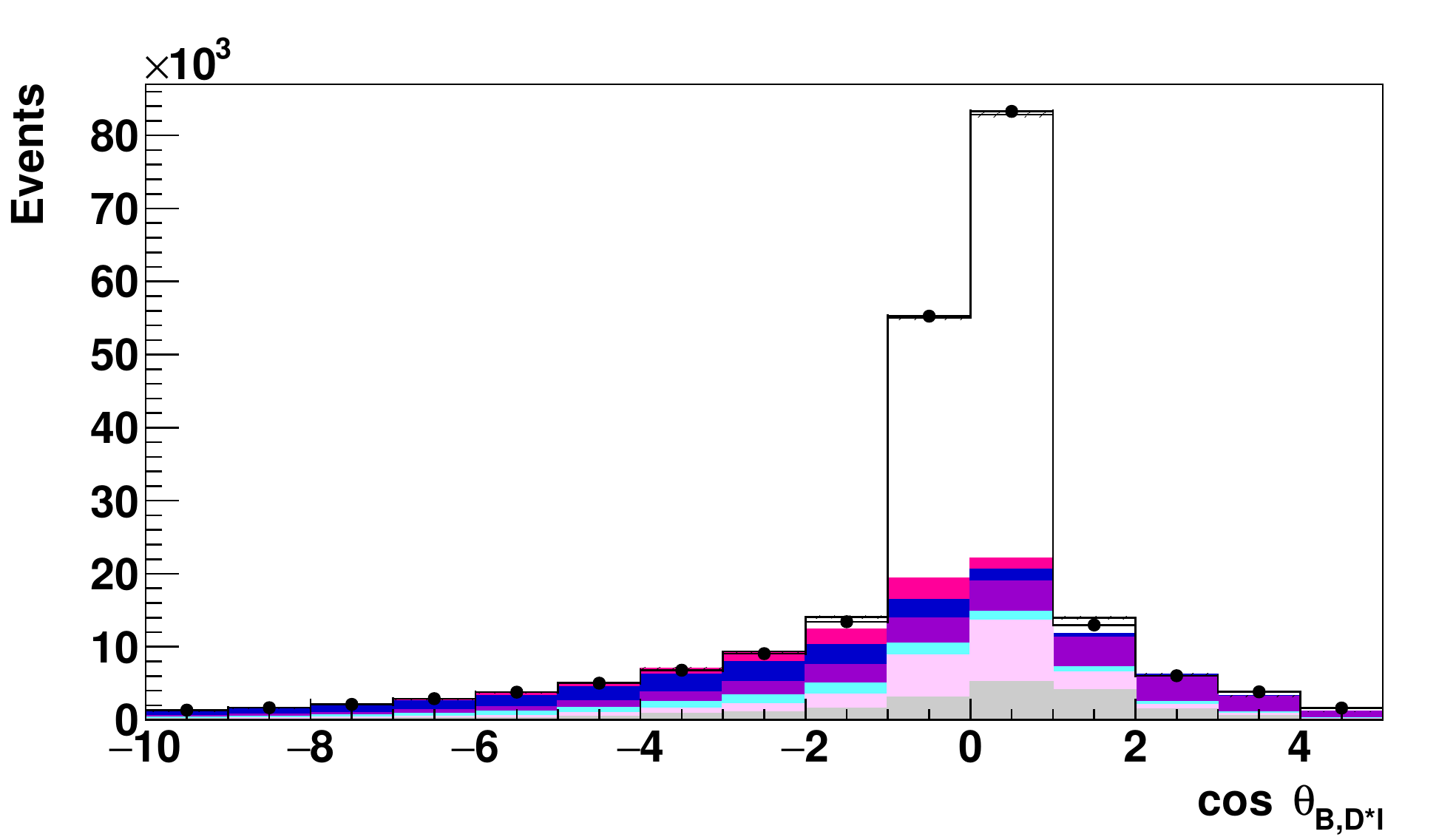} 
 \caption{Result of the fits to the ($\cos\theta_{B,D*\ell}$, $\Delta M$, $p_\ell$) distributions in the $e$ mode (left) and $\mu$ mode (right). The bin boundaries are discussed in the text. The points are on-resonance data, where the uncertainties are smaller than the markers." . The colour scheme is defined in the figure.}
 \label{fig:bkgfit}
 \end{figure*}

To estimate the background yields we perform a binned maximum log likelihood fit of the $D^*\ell$ candidates in three variables, $\Delta$M, $\cos\theta_{B,D^*\ell}$, and $p_\ell$. The bin ranges are as follows:
\begin{itemize}
\item $\Delta$M: 5 equidistant bins in the range $[0.141,~0.156]$ GeV/$c^2$. 
\item $\cos\theta_{B,D^*\ell}$: 15 equidistant bins  in the range $[-10,~5]$. 
\item $p_\ell$: 2 bins in the ranges $[0.6,~0.85,~3.0]$ GeV/$c$ for muons and $[0.3, 0.80, 3.0]$  GeV/$c$ for electrons.
\end{itemize}
Prior to the fit, the residual continuum background is estimated from off-resonance data and scaled by the off-on resonance integrated luminosities and the 1/$s$ dependence of the $e^+e^-\to q \bar q$ cross section. The kinematics of the off and on-resonant continuum background is expected to be slightly different and therefore binned correction weights are determined using MC and applied to the scaled off-resonance data. The remaining background components are modelled with MC simulation after correcting for the most recent decay modelling parameters, and for differences in reconstruction efficiencies between data and MC. Corrections are applied to the lepton identification efficiencies, hadron misidentification rates, and slow pion tracking efficiencies. The data/MC ratios for high momentum tracking efficiencies are consistent with unity and are only considered in the systematic uncertainty estimates.
The results from the background fits are in Table~\ref{tab:bkgfit} and Fig.~\ref{fig:bkgfit}.

After applying all analysis criteria and subtracting background, a total of 90738 and 89082 $B^{0} \to D^{*-} e^+ \nu_e$ and $B^{0} \to D^{*-} \mu^+ \nu_\mu$ signal decays are found respectively.

\begin{table*}[htb]
\centering
\caption{Signal and background fractions ($\%$) for events selected in the signal region of  ($|\cos\theta_{B,D^{*}\ell}|<1$, $0.144~\rm GeV/c^2<\Delta M<0.147~\rm GeV/c^2$, $p_{e}>0.80~\rm GeV/c$, $p_{\mu}>0.85~\rm GeV/c$).}
 \renewcommand{\arraystretch}{1.3}

\begin{tabularx}{0.8\linewidth}{lYYYY}
\hline \hline
             		 & \textbf{SVD1(e)} 			&\textbf{SVD1($\mu$)}    &\textbf{SVD2 (e)} &\textbf{SVD2 ($\mu$)} \\ \hline
Signal yield   & 19318    		& 19748                           &  88622                      &  87060  \\ 
\hline
Signal ~~~~~~~  &  \small 79.89 $\pm$ 0.58  & 80.12  $\pm$ 0.52    & 81.00  $\pm$ 0.19   &  79.86 $\pm$ 0.20 \\ 
Fake $\ell$      &  \small 0.09 $\pm$  0.16       & 1.55  $\pm$ 0.69      &  0.10     $\pm$ 0.79    & 1.15  $\pm$ 0.38 \\ 
Fake $D^{*}$   &  \small 3.05 $\pm$  0.09      &  2.89  $\pm$ 0.06     & 2.94    $\pm$ 0.01    & 2.81  $\pm$ 0.01\\ 
$D^{**}$          &  \small 5.82 $\pm$   0.40     & 4.00  $\pm$ 0.24   & 5.08 $\pm$ 0.14       &  3.62  $\pm$  0.08 \\ 
Signal corr.     & \small 1.24 $\pm$  0.34       &  1.99 $\pm$ 0.38   & 1.42  $\pm$ 0.07      &  2.39 $ \pm$ 0.14  \\ 
Uncorrelated    & \small  5.81 $\pm$  0.50    &  5.01 $\pm$ 0.58   & 4.96  $\pm$ 0.15      &  5.00 $\pm$ 0.24 \\ 
Continuum    & \small  4.11 $\pm$  0.64       &  4.44 $\pm$ 0.74   & 4.48     $\pm$ 0.38       &  5.16 $\pm$ 0.46 \\ 
\hline
\end{tabularx}
\label{tab:bkgfit}
\end{table*}

\section{Measurement of differential distributions}\label{sec-theory}
Measurement of the decay kinematics requires good knowledge of the signal $B$ direction to constrain the neutrino momentum 4-vector. To determine the $B$ direction we estimate the CM frame momentum vector of the non-signal $B$ meson by summing the momenta of the remaining particles in the event ($\vec p_{\rm incl.}^*$) and choose the direction on the cone that minimises the difference to $-p_{\rm incl.}^*$. To determine $p_{\rm incl.}^*$ we exclude tracks that do not pass near the interaction point. The impact parameter requirements depend on the transverse momentum of the track, $p_{\rm T}$, and are set to:
\begin{itemize}
\item $p_{\rm T}<250$ MeV$/c$: $dr<20$ cm, $|dz| <100$ cm,
\item $p_{\rm T}<500$ MeV$/c$: $dr<15$ cm, $|dz| <50$ cm,
\item $p_{\rm T}\ge 500$ MeV$/c$: $dr<10$ cm, $|dz| <20$ cm.
\end{itemize}
Some track candidates may be counted multiple times, due to low momentum particles spiralling in the CDC, or due to fake tracks fit to a similar set of detector hits as the real track. These are removed by looking for pairs of tracks with similar kinematics, travelling in the same direction with the same electric charge, or in the opposite direction with the opposite electric charge. Isolated clusters that are not matched to the signal particles (i.e. from photons or $\pi^0$ decays) are required to have lower energy thresholds to mitigate beam induced background, and are 50, 100 and 150 MeV in the barrel, forward and backward end-cap regions, respectively. We compute $\vec p_{\rm incl.}$ by summing the 3-momenta of the selected particles:
\begin{eqnarray}
\vec p_{\rm incl.} = \sum_i \vec p_i~,
\end{eqnarray}
where the index $i$ denotes all isolated clusters and tracks that pass the above criteria. This vector is then translated into the CM frame. The energy component, $E_{\rm incl.}^*$, is set to the experiment dependent beam energies through $E_{\rm beam}^*=\sqrt{s}/2$.

We find that the resolutions of the kinematic variables are 0.020 for $w$, 0.038 for $\cos\theta_\ell$, 0.044 for $\cos\theta_{\rm v}$ and 0.210 for $\chi$. Based on these resolutions, and the available data sample, we split each distribution into 10 equidistant bins for the $|V_{cb}|$ and form factor fits.

\subsection{Fit to the CLN Parameterization}
We perform a binned $\chi^2$ fit to determine the following quantities in the CLN parameterization: the product ${\mathcal F_1} |V_{cb}|$, and the three parameters $\rho^2$, $R_1(1)$ and $R_2(1)$ that parameterise the form factors.  We use a set of one-dimensional projections of $w$, $\cos\theta_\ell$, $\cos\theta_{\rm v}$ and $\chi$. This reduces complications in the description of the six background components and their correlations across four dimensions. This approach introduces finite bin-to-bin correlations that must be accounted for in the $\chi^2$ calculation.

We choose equidistant binning in each kinematic observable, as described above, and set the ranges according to their kinematically allowed limits. The exception is $w$: while the kinematically allowed range is between 1 and 1.504, we restrict this to between 1 and 1.50 such that we can ignore the finite mass of the lepton in the interaction.

The number of expected signal events produced in a given bin  $i$, $N_i^{\rm prod.}$, is given by
\begin{eqnarray}
N_{i}^{\rm prod.} &=& N_{B^0} {\cal B}(D^{*+}\to D^0\pi^+) \nonumber\\
                   & &\times{\cal B}(D^0\to K^- \pi^+)\tau_{B^0}\Gamma_i~,
\end{eqnarray}
where $N_{B^0}$ is the number of $B^0$ mesons in the data sample, ${\cal B}(D^{*+}\to D^0\pi^+) $ and ${\cal B}(D^0\to K^- \pi^+)$ are the $D^*$ and $D^{0}$ branching ratios into the final state studied in this analysis, $\tau_{B^0}$ is the $B^0$ lifetime, and $\Gamma_i$ is the width obtained by integrating the CLN theoretical expectation within the corresponding bin boundaries. The values of the $D^*$ and the $D^{0}$ branching fractions as well as the $B^0$ lifetime are taken from the PDG.  The value of $N_{B^{0}}$ is calculated using $N_{B^{0}} = 2 \times f_{00} \times N_{BB}$ where $N_{BB}$ is stated in Section~\ref{data-section} and $f_{00} = 0.486 \pm 0.006$~\cite{cite-hflav}.
The expected number of events, $N_i^{\rm exp.}$, must take into account finite detector resolution and efficiency, 
\begin{eqnarray}
N_{i}^{\rm exp} = \sum_{j=1}^{40} (R_{ij}\epsilon_jN_j^{\rm prod}) + N^{\rm bkg}_i~,
\end{eqnarray}
where $\epsilon_j$ is the probability that an event generated in bin $j$ is reconstructed and passes the analysis selection criteria, and $R_{ij}$ is the detector response matrix (the probability that an event generated in bin $j$ is observed in bin $i$). $N_{i}^{\rm bkg}$ is the number of expected background events as constrained from the total background yield fit.

\begin{figure*}[htb]
 \centering
 \includegraphics[width=0.45\linewidth]{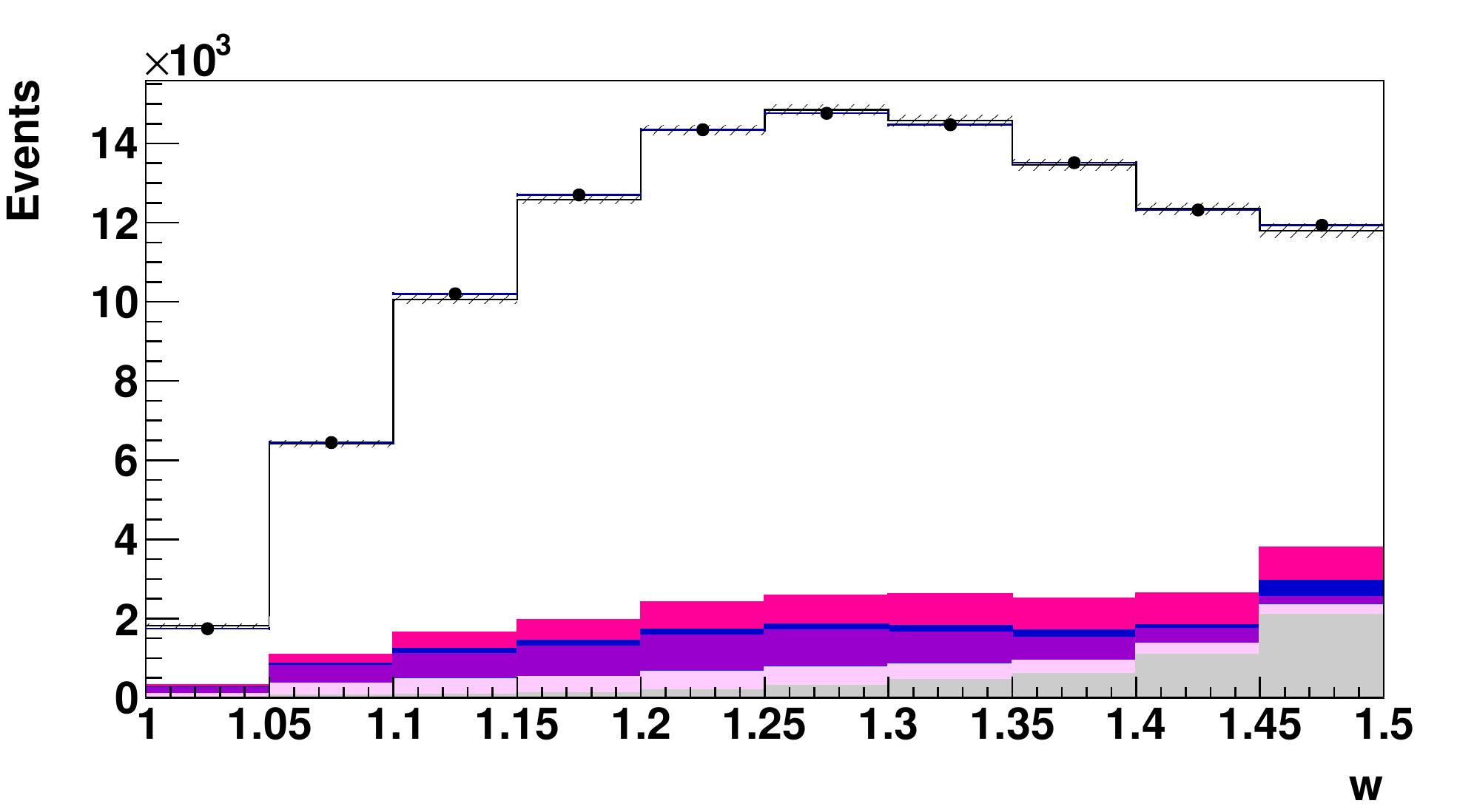} 
 \includegraphics[width=0.45\linewidth]{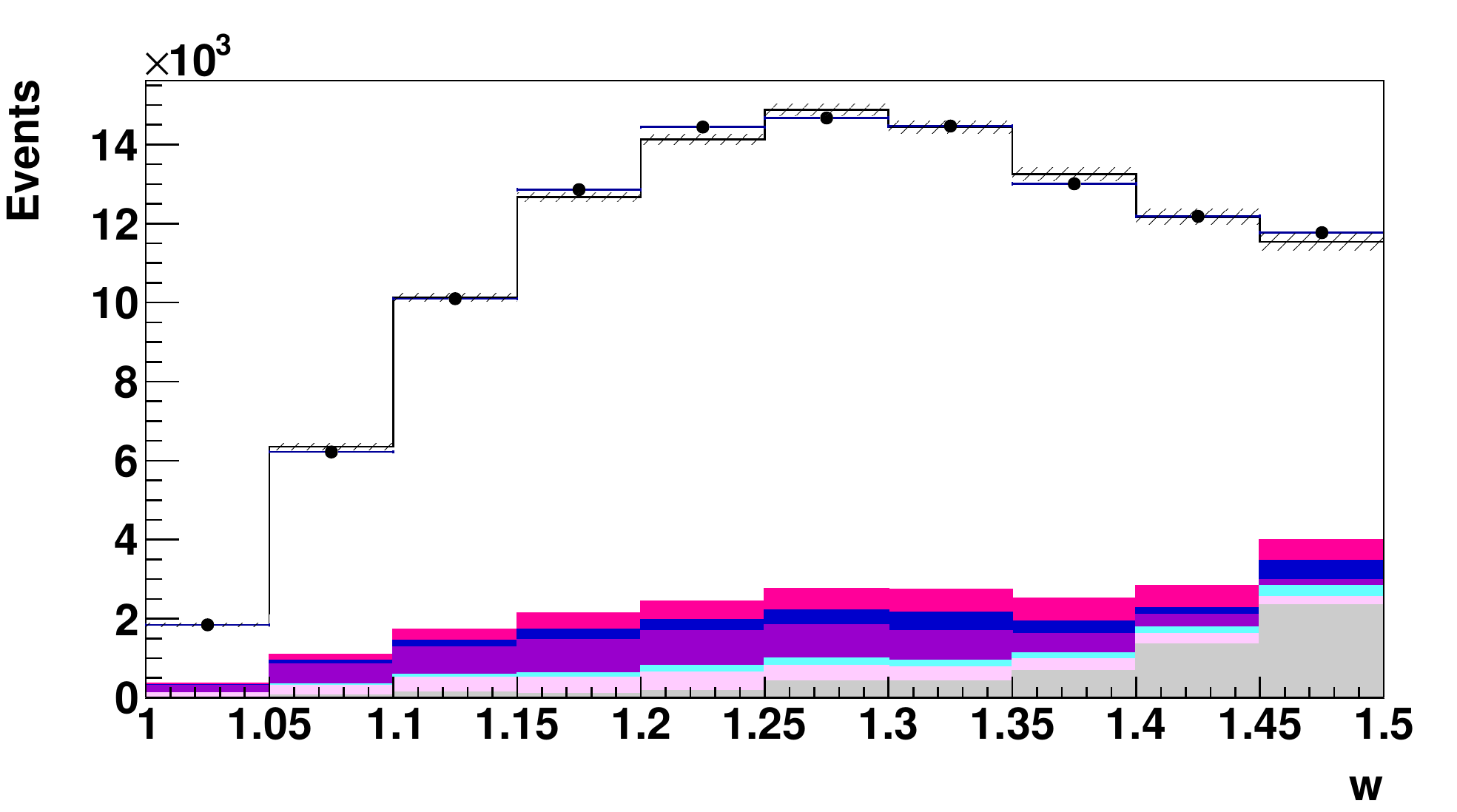} 
  \includegraphics[width=0.45\linewidth]{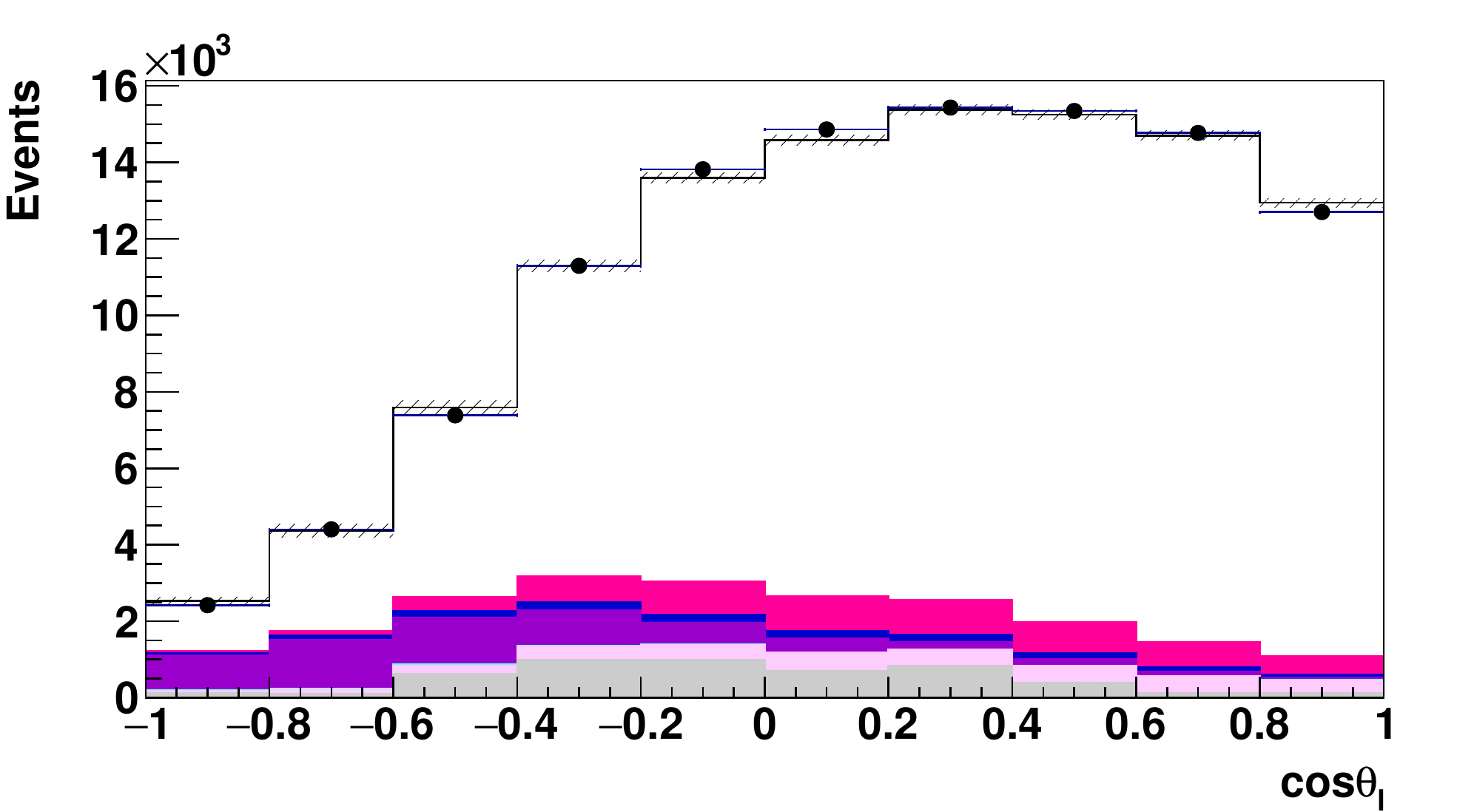} 
  \includegraphics[width=0.45\linewidth]{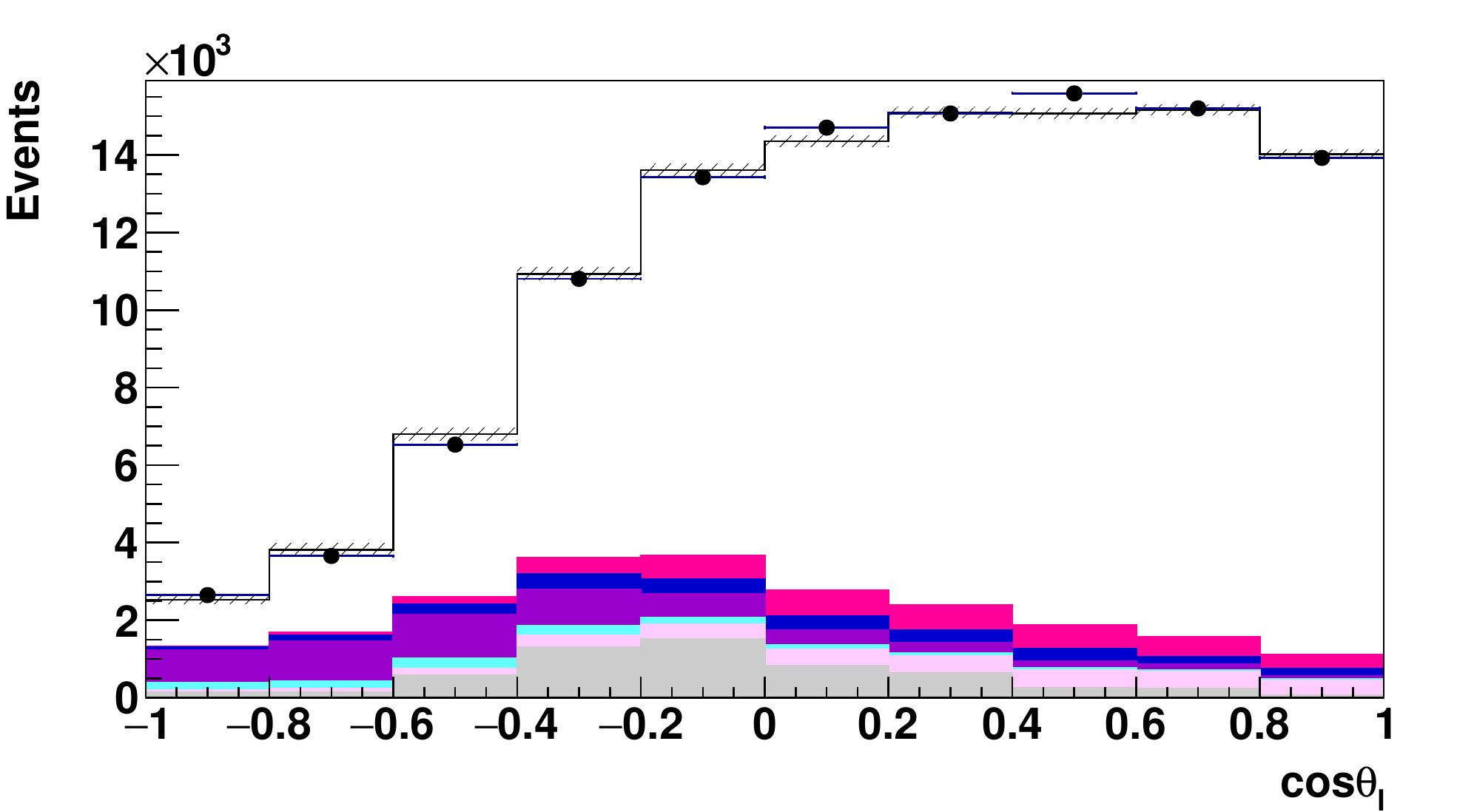} 
  \includegraphics[width=0.45\linewidth]{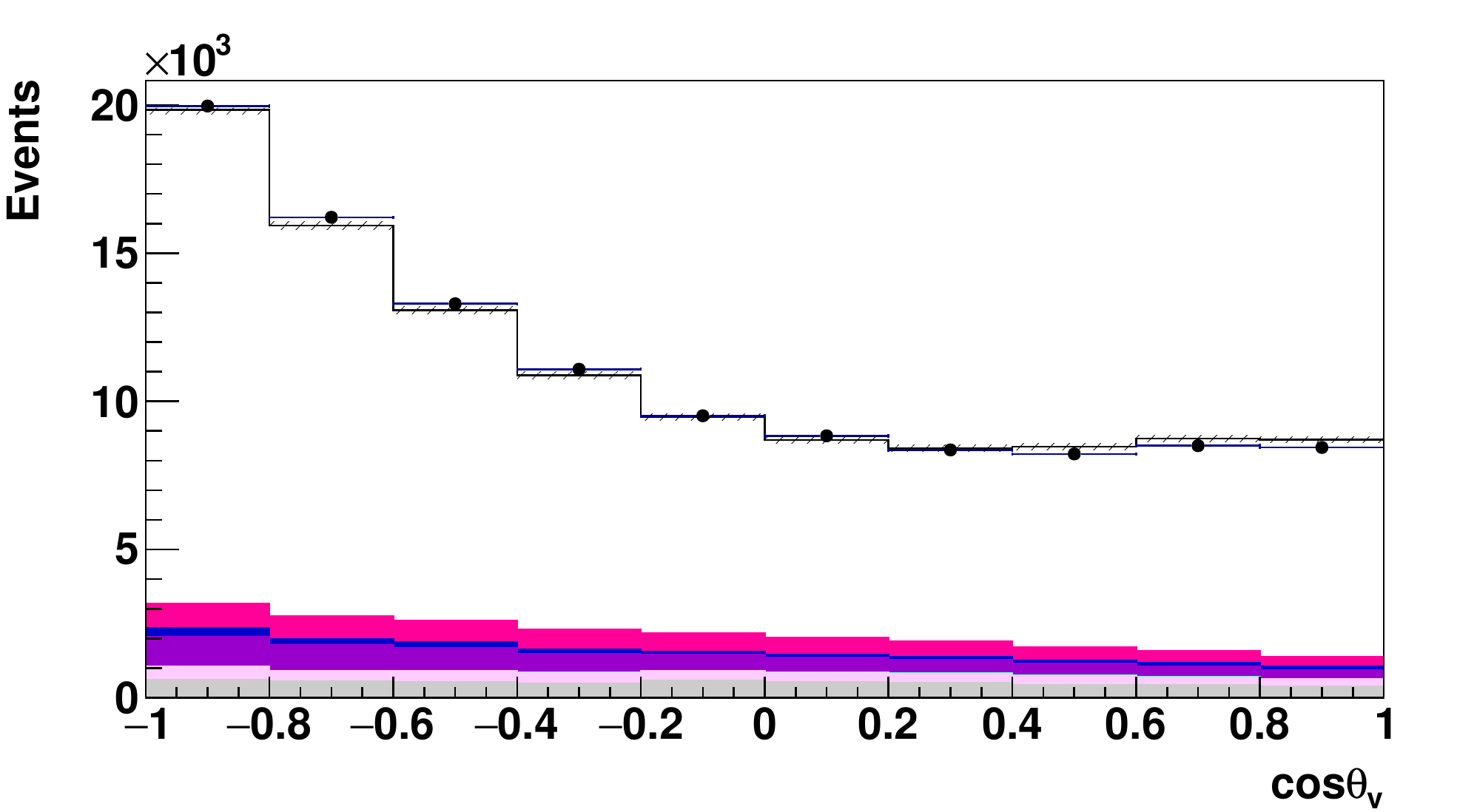} 
  \includegraphics[width=0.45\linewidth]{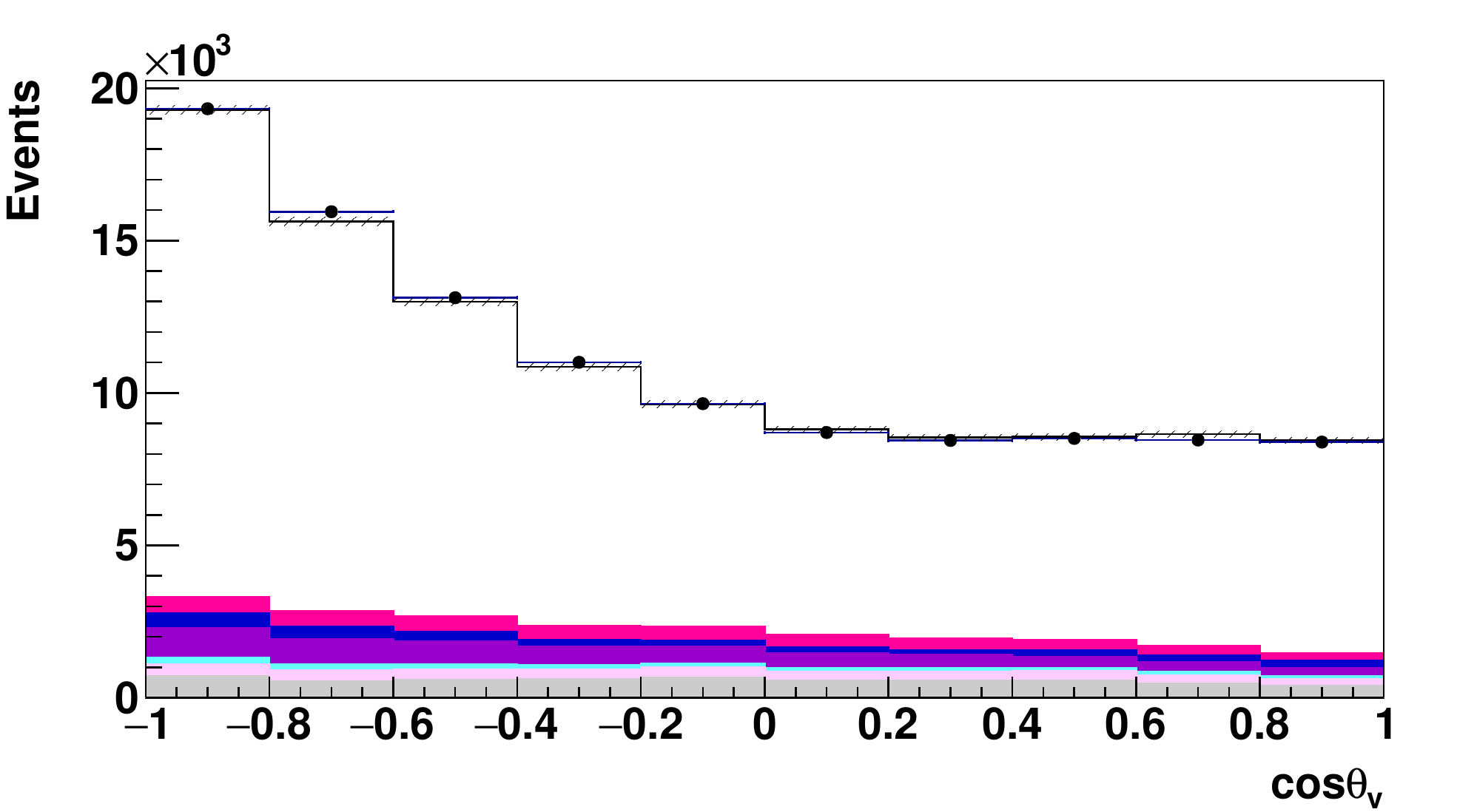} 
  \includegraphics[width=0.45\linewidth]{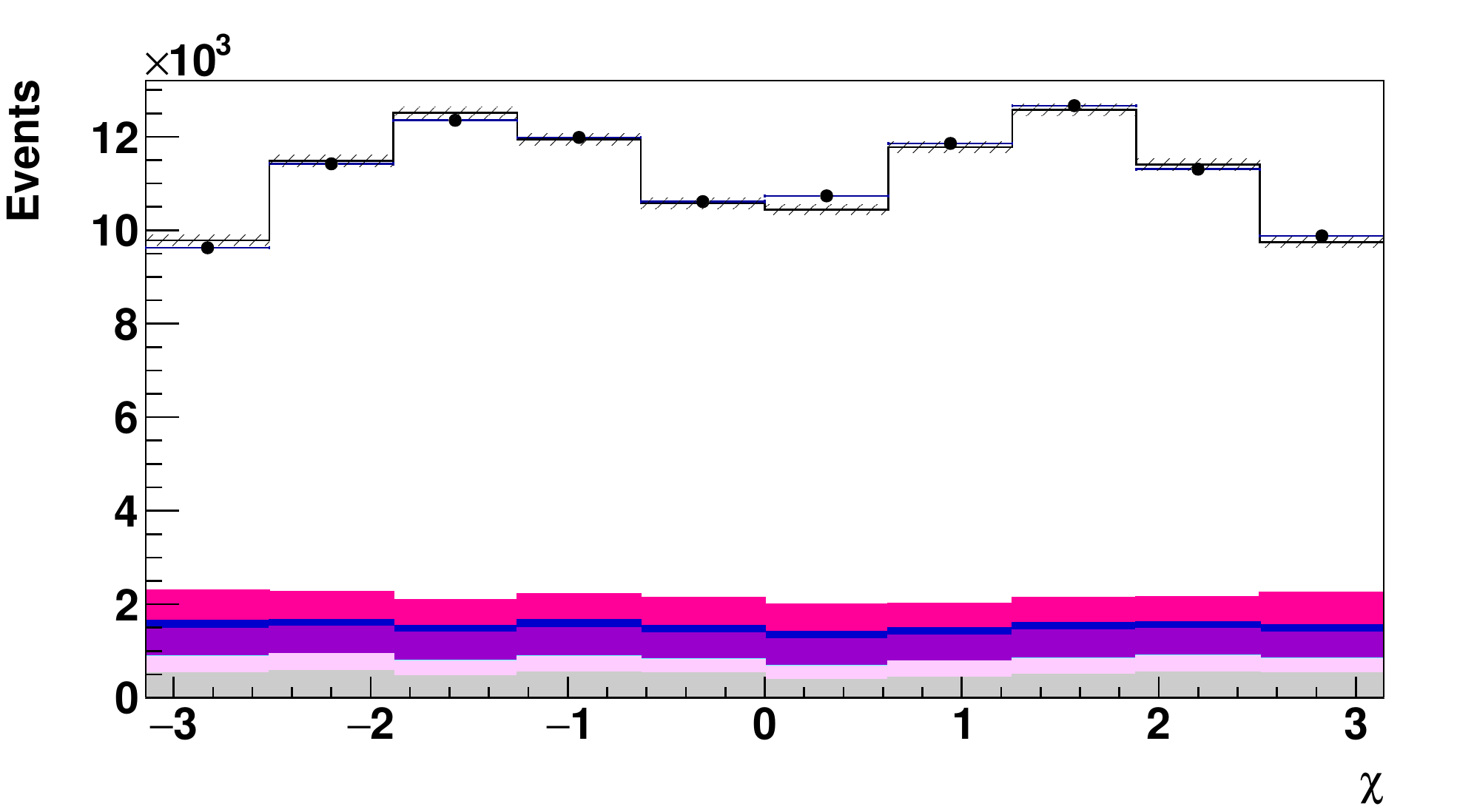} 
  \includegraphics[width=0.45\linewidth]{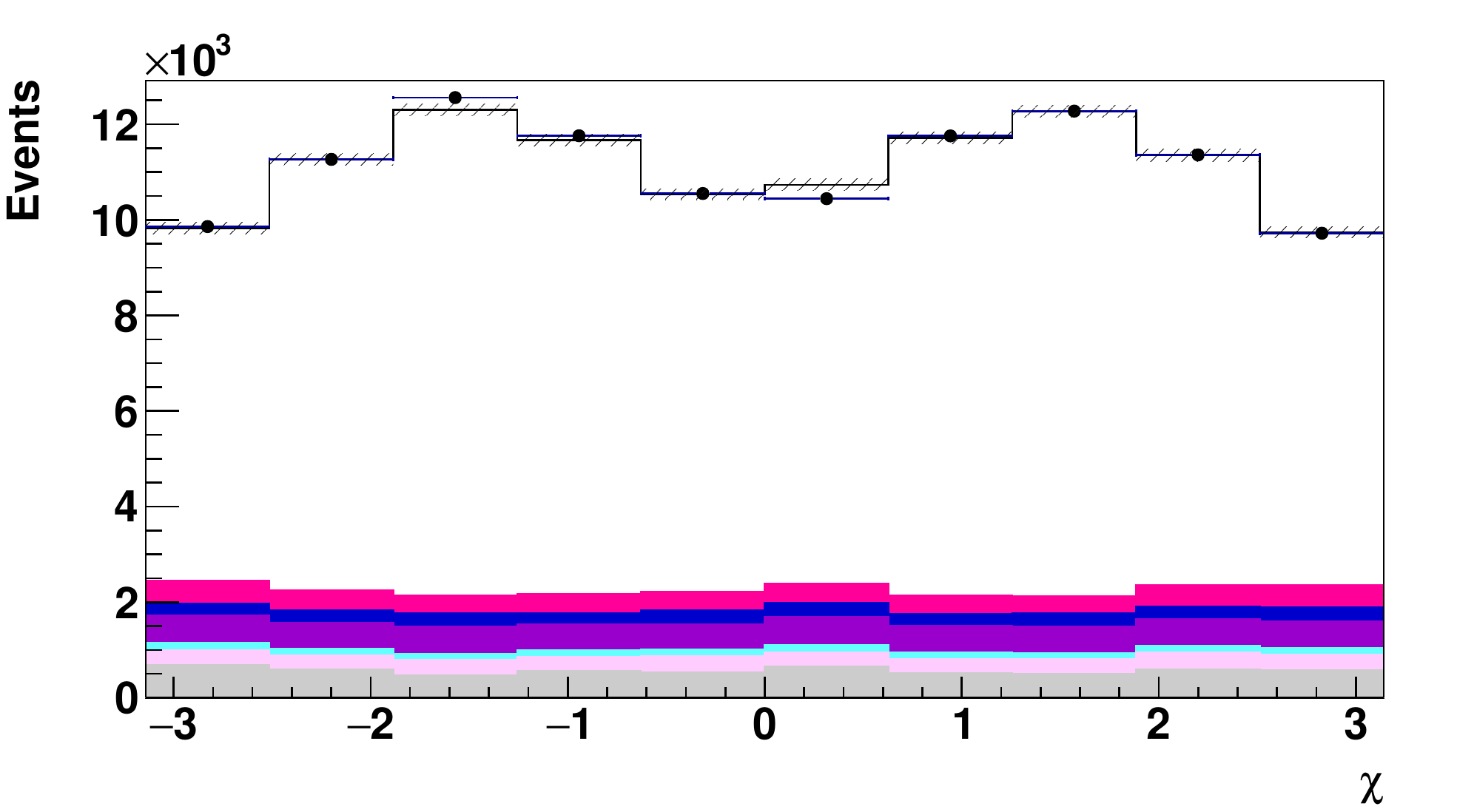} 
\caption{Results of the fit with the CLN form factor parameterization. The results from the SVD1 and SVD2 samples are added together. The electron modes are on the left and muon modes on the right.  The points with error bars are the on-resonance data. Where not shown, the uncertainties are smaller than the black markers. The histograms are, from top to bottom, the signal component, $B \to D^{**}$ background, signal correlated background, uncorrelated background, fake $\ell$ component, fake $D^*$ component and continuum.}
 \label{fig:clnfit}
 \end{figure*}

In the nominal fit we use the following $\chi^2$ function based on a forward folding approach:
\begin{eqnarray}\label{chi2}
\chi^2 = \sum_{i,j}^{} \left(N_i^{\rm obs} - N_i^{\rm exp} \right)C^{-1}_{ij} \left( N_j^{\rm obs} - N_j^{\rm exp} \right),
\end{eqnarray}
where $N_i^{\rm obs}$ are the number of events observed in bin $i$ of our data sample, and $C_{ij}^{-1}$ is the inverse of the covariance matrix C. The covariance matrix is the variance-covariance matrix whose diagonal elements are the variances and the off-diagonal elements are the covariance of the elements from the $i^{\rm th}$ and $j^{\rm th}$ positions. The covariance is calculated for each pair of bins in either $w$, $\cos\theta_\ell$, $\cos\theta_V$ and $\chi$. The off-diagonal elements are calculated as,
\begin{eqnarray}
C_{ij} = N \mathcal{P}_{ij} - N\mathcal{P}_i\mathcal{P}_j, ~~~ \forall i\neq j~,
\end{eqnarray}
where $\mathcal{P}_{ij}$ is the relative probability of the two-dimensional histograms between observable pairs,  $\mathcal{P}_i$ and $\mathcal{P}_j$ are the relative probabilities of the one-dimensional histograms of each observable, and $N$ is the total size of the sample. The diagonal elements are the variances of $N_i^{\rm exp}$ and are calculated as,
\begin{eqnarray}
\sigma_i^2 &=& \sum_{j=1}^{40} \left[ R_{ij}^2 \epsilon_j^2N_{j}^{\rm th} + R_{ij}^2 \frac{1-\epsilon_j}{N_{\rm data}}(N^{\rm th}_j)^2 \right.\nonumber \\
&& + R_{ij}\frac{1-R_{ij}}{N^\prime_{\rm data}}\epsilon_j^2(N^{\rm th})^2 + R_{ij}^2\frac{1-\epsilon_j}{N_{\rm MC}} \nonumber \\
&& \left. + R_{ij} \frac{1-R_{ij}}{N^\prime_{\rm MC}} \epsilon_j^2(N^{\rm th}_j)^2 \right] + \sigma^2(N_i^{\rm bkg}).
\end{eqnarray} 
which uses the Poisson uncertainty associated with the number of events in the MC and data in each bin, and the final term is the total error associated with the background arising from the background fit procedure.
We have tested this fit procedure using MC simulated data samples and all results are consistent with expectations, showing no signs of bias.
The results from the fit are summarized in Table~\ref{tab:clnfit} and the fit correlation coefficients are given in Table~\ref{tab:clnfitcorr}. The comparison between data and the form factor fit is shown in Fig.~\ref{fig:clnfit}.   

\begin{table*}[htb]
\caption{Fit results for the four sub-samples in the CLN parameterization  where the following parameters are floated: $\rho^{2}$, $R_{1}(1)$, $R_{2}(1)$ along with $\mathcal F(1)|V_{cb}\eta_{EW}|$. The $p$-value corresponds to the $\chi^2/$ndf using the statistical errors only.} 
 \renewcommand{\arraystretch}{1.3}
\begin{tabularx}{0.8\linewidth}{l  YYYY } 
\hline \hline
        &  SVD1 $e$ & SVD1 $\mu$ &   SVD2 $e$    & SVD2 $\mu$ \\ 
\hline 
 $\rho^{2}$ & 1.165 $\pm$ 0.099    & 1.165 $\pm$ 0.102  & 1.087 $\pm$ 0.046  & 1.095 $\pm$ 0.051\\ 
$R_{1}(1)$  & 1.326 $\pm$ 0.106    & 1.336 $\pm$ 0.103    & 1.117 $\pm$ 0.040  & 1.287 $\pm$ 0.047\\ 
$R_{2}(1)$  & 0.767 $\pm$ 0.073    & 0.777 $\pm$ 0.074    & 0.861 $\pm$ 0.030  & 0.884 $\pm$ 0.034\\ 
$\mathcal F(1)|V_{cb}|\eta_{\rm EW}\times10^{3}$  & 34.66 $\pm$ 0.48 & 35.01 $\pm$ 0.50   & 35.25 $\pm$ 0.23 & 34.98 $\pm$ 0.25\\ 
\hline
$\chi^{2}$/ndf  &    35/36             & 36/36  &   44/36           &   43/36    \\ 
$p$-value &   0.52       &    0.47 &  0.17     &   0.20    \\ 
$\mathcal{B}(B^{0} \to D^{*-} \ell^+ \nu_\ell)$ [\%] &4.89 $\pm$ 0.06   &4.96 $\pm$ 0.06    &4.93 $\pm$ 0.03     &  4.86 $\pm$ 0.03  \\ \hline
\end{tabularx}
\label{tab:clnfit}
\end{table*}

\begin{table}[htb]
\centering
\caption{Statistical correlation matrix of the fit to the full sample in the CLN parameterization.}
 \renewcommand{\arraystretch}{1.3}
\begin{tabularx}{0.9\linewidth}{l  YYYY} 
\hline 
\hline 
\textbf{ } 	& \textbf{$\rho^{2}$} 	& $R_{1}(1)$ & $R_{2}(1)$ & $\mathcal F(1)|V_{cb}|$ \\ \hline
\textbf{ $\rho^{2}$} 	            & $+1.000$	& $+0.593$ &$-0.883$ & $+0.655$\\
\textbf{ $R_{1}(1)$} 	            & 	    & $+1.000$ &$-0.692$ &$-0.062$\\ 
\textbf{ $R_{2}(1)$} 	            &  	    &       &+1.000 &$-0.268$\\
\textbf{ $\mathcal F(1)|V_{cb}|$} 	&  	    &       &       & $+1.000$\\ 
\hline
\end{tabularx}
\label{tab:clnfitcorr}
\end{table}

\subsection{Branching Fraction Measurement}
The branching fraction for $\mathcal{B}(B^{0} \to D^{*-} \ell^+ \nu_\ell$) is obtained with the relation,
\begin{equation}
 \mathcal{B}=\frac{N_{\rm signals}}{\epsilon \times {\cal B}(D^{*+}\to D^0\pi^+)  \times {\cal B}(D^0\to K^- \pi^+)}~,
\end{equation}
where $N_{\rm signals}$ are signals after applying all the selection criteria, $\epsilon$ is the efficiency of the decay, while the values of the branching fractions ${\cal B}(D^{*+}\to D^0\pi^+)$ and ${\cal B}(D^0\to K^- \pi^+)$ are taken from PDG. The branching fraction is calculated for all the samples separately, as well as combined. 

\subsection{Fit to the BGL Parameterization}
\begin{figure*}[htb]
 \centering
\includegraphics[width=0.45\linewidth,page=1]{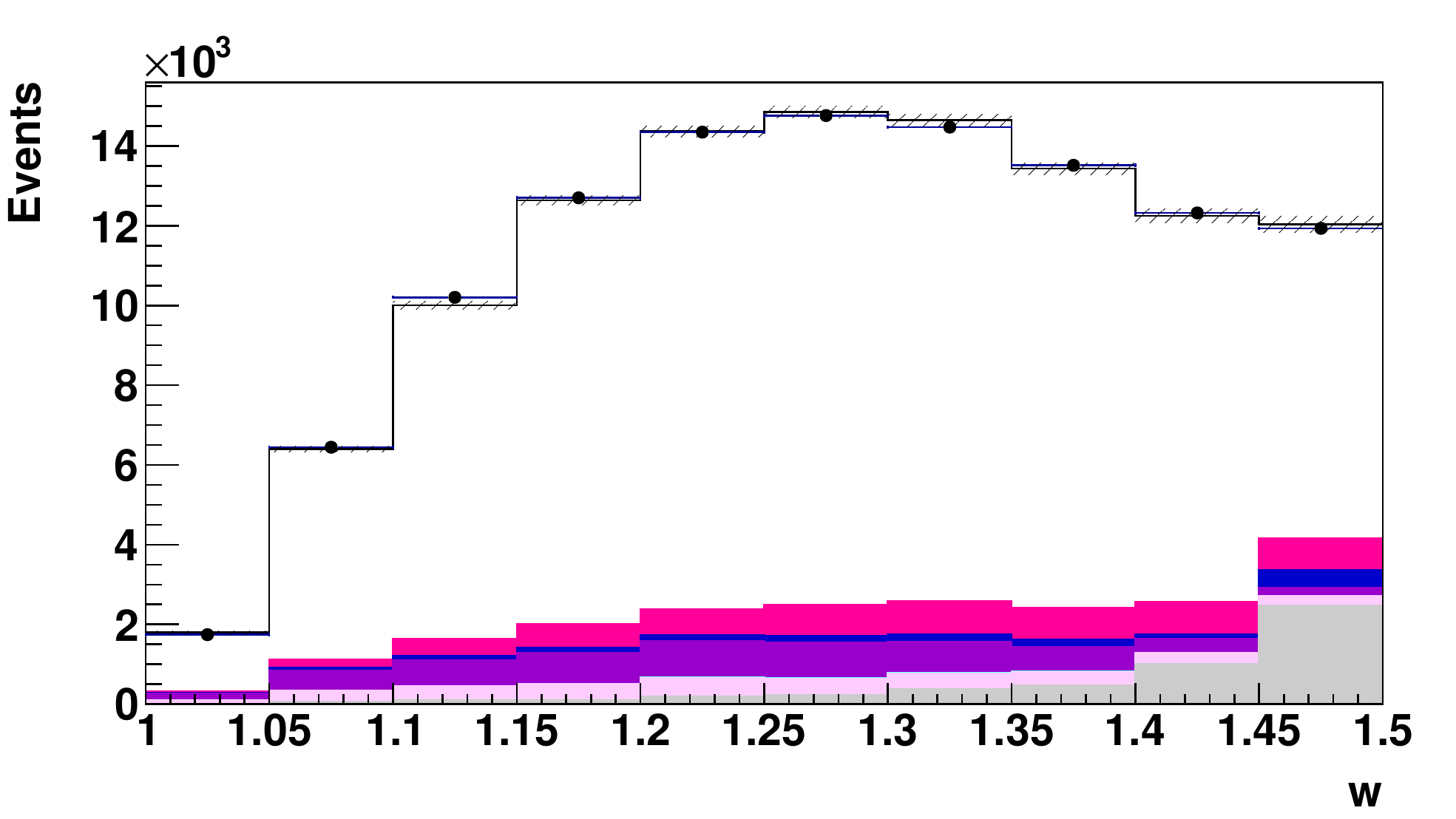} 
 \includegraphics[width=0.45\linewidth,page=1]{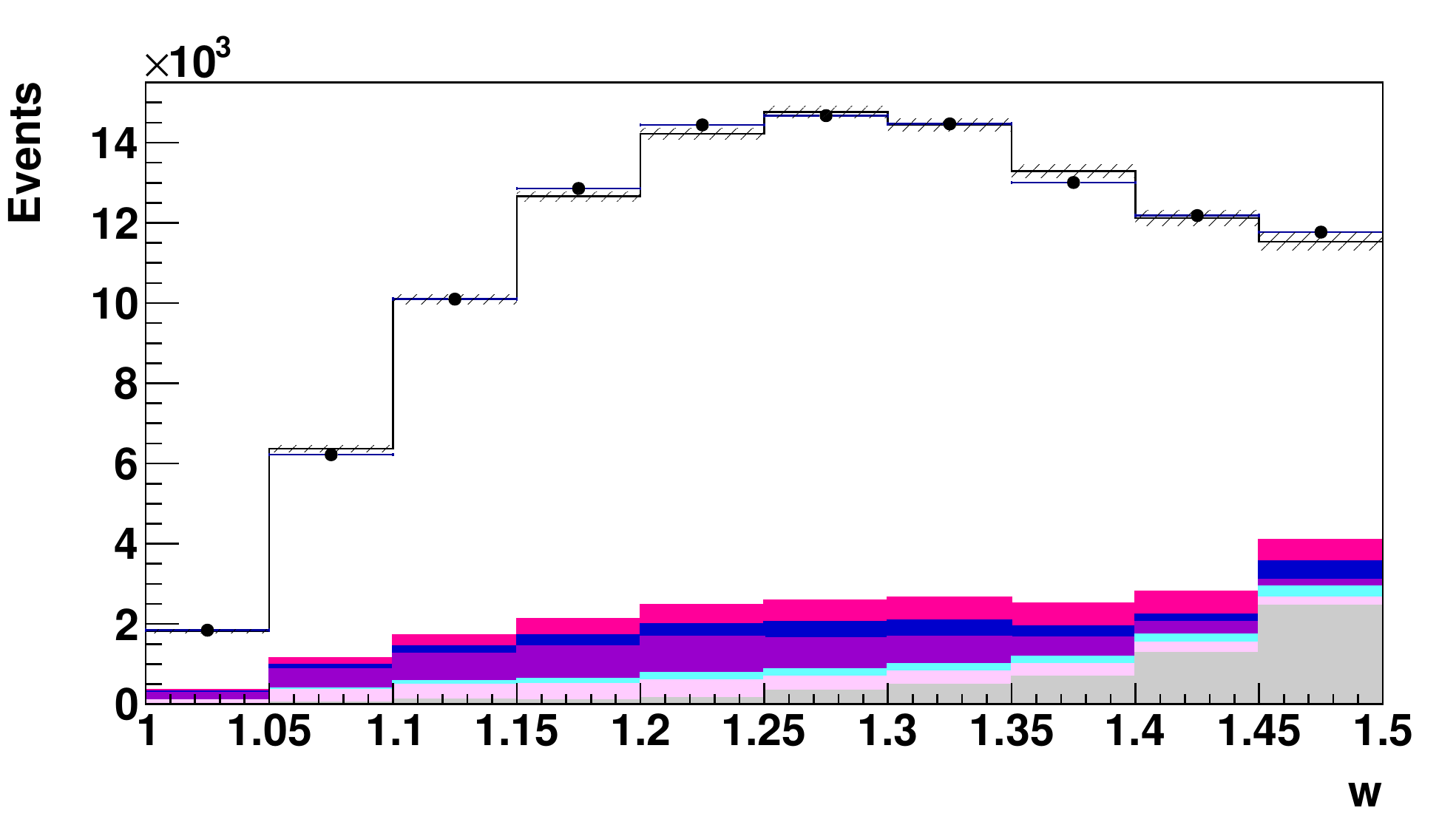} 
  \includegraphics[width=0.45\linewidth,page=4]{bgl_plot_svd0_lepgen0.pdf} 
  \includegraphics[width=0.45\linewidth,page=4]{bgl_plot_svd0_lepgen1.pdf} 
  \includegraphics[width=0.45\linewidth,page=3]{bgl_plot_svd0_lepgen0.pdf} 
  \includegraphics[width=0.45\linewidth,page=3]{bgl_plot_svd0_lepgen1.pdf} 
  \includegraphics[width=0.45\linewidth,page=2]{bgl_plot_svd0_lepgen0.pdf} 
  \includegraphics[width=0.45\linewidth,page=2]{bgl_plot_svd0_lepgen1.pdf} 
 \caption{Results of the fit with the BGL form factor parameterization. The results from the SVD1 and SVD2 samples are added together. The electron modes are on the left and muon modes on the right.  The points with error bars are the on-resonance data. Where not shown, the uncertainties are smaller than the black markers. The histograms are, top to bottom, the signal component, $B \to D^{**}$ background, signal correlated background, uncorrelated background, fake $\ell$ component, fake $D^*$ component and continuum.}
 \label{fig:bglfit}
 \end{figure*}

To perform the fit to the BGL parameterization we follow the approach in Ref.~\cite{cite-grinstein}.  We truncate the series in the expansion for $a^f$ and $a^g$ terms at $\mathcal O(z^2)$ and order $\mathcal O(z^3)$ for $a^{\mathcal{F}_1}$. Due to very large correlations when introducing $a^g_1$ we remove it from the nominal fit results.
This results in five free parameters (one more than in the CLN fit), defined as $\tilde a_i^f=|V_{cb}|\eta_{\rm EW}a_i^f$ where $i=0,1$, $\tilde a_i^g=|V_{cb}|\eta_{\rm EW}a_i^g$ where $i=1$ and $\tilde a_i^{F_1}=|V_{cb}|\eta_{\rm EW}a_i^{F_1}$, where $i=1,2$. This number of free parameters can describe the data well, while higher order terms will not be well constrained unless additional information from lattice QCD is introduced. We perform a $\chi^2$ fit to the data with the same procedure as for the CLN fit described above. The resulting value for $|V_{cb}|$ is consistent with that from the CLN parameterization. The fit results are given in Table~\ref{tab:bglfit2} and Fig.~\ref{fig:bglfit}. The linear statistical correlation coefficients are listed in Table~\ref{tab:bglcorrsum}. Correlations can be high in this fit approach: only the SVD1+SVD2 combined samples are fitted as the fit does not converge well with the smaller SVD1 data set.

\begin{table*}[htb]
 \centering
 \caption{Fit results for the electron and muon sub-samples in the BGL parameterization where the following parameters are floated: $\tilde a_0^f$, $\tilde a_1^f$, $\tilde a_1^{F_1}$, $\tilde a_2^{F_1}$, $\tilde a_0^g$ along with $\mathcal F(1)|V_{cb}|\eta_{\rm EW}$ (derived from $\tilde a_0^f$). The $p$-value corresponds to the $\chi^2/$ndf using the statistical errors only.}
\label{tab:bglfit2}
\setlength{\tabcolsep}{12pt}
 \renewcommand{\arraystretch}{1.3}
\begin{tabularx}{0.65\linewidth}{lYY} 
\hline \hline
&  $e$ & $\mu$ \\ 
\hline 
$\tilde a_0^f\times10^{2}$      &$-0.0507$  $\pm$ 0.0005  & $-0.0505$  $\pm$ 0.0006  \\
$\tilde a_1^f\times10^{2}$      &$-0.0673$  $\pm$ 0.0220  & $-0.0626 $ $\pm$ 0.0252   \\
$\tilde a_1^{F_1}\times10^{2}$  &$-0.0292$  $\pm$ 0.0086  &$ -0.0247$  $\pm$ 0.0096   \\
$\tilde a_2^{F_1}\times10^{2}$  &$+0.3407$  $\pm$ 0.1674  & $+0.3123$  $\pm$ 0.1871   \\
$\tilde a_0^g\times10^{2}$      &$-0.0864$  $\pm$ 0.0024  & $-0.0994$  $\pm$ 0.0027  \\
\hline
$\mathcal F(1)|V_{cb}|\eta_{\rm EW}\times10^{3}$ & 35.01 $\pm$ 0.31 & 34.84 $\pm$ 0.35  \\ 
\hline
$\chi^{2}$/ndf  &    48/35             & 43/35   \\ 
$p$-value &   0.08    &    0.26  \\ 
$\mathcal{B}(B^{0} \to D^{*-} \ell^+ \nu_\ell)$ [\%] &4.91 $\pm$ 0.02 & 4.88 $\pm$ 0.03 \\
\hline
\end{tabularx}
\end{table*}

\begin{table}[htb]
\centering
\caption{Statistical correlation matrix of the fit to the full sample in the BGL parameterization.}
\label{tab:bglcorrsum}
 \renewcommand{\arraystretch}{1.3}
\begin{tabularx}{0.9\linewidth}{lYYYYY} 
\hline 
\textbf{ } 	& \textbf{$\tilde a_0^f$ } 	& $\tilde a_1^f$ & $\tilde a_1^F$ & $\tilde a_2^F$ & $\tilde a_0^g$  \\ \hline\hline
\textbf{$\tilde a_0^f$} & $+1.000$	& $-0.790$ & $-0.775$  & $+0.669$ & $-0.038$  \\
\textbf{$\tilde a_1^f$} &       &  $+1.000$ & $+0.472$  & $-0.411$ & $-0.406$  \\ 
\textbf{$\tilde a_1^F$} &   	&        & $+1.000$   & $-0.981$ & $+0.071$  \\
\textbf{$\tilde a_2^F$} &   	&        &         & $+1.000$ & $-0.057$ \\
\textbf{$\tilde a_0^g$} &   	&        &         &        & $+1.000$  \\
\hline
\end{tabularx}
\end{table}
\section{Systematic Uncertainties}\label{sec-syst}
To estimate systematic uncertainties on the partial branching fractions, form factor parameters, and $|V_{cb}|$, we consider the following sources: background component normalizations, tracking efficiency, charm branching fractions, $B\to D^{**}\ell \nu$ branching fractions and form factors, the $B^0$ lifetime, and the number of $B^0$ mesons in the data sample. The systematic uncertainties on the branching fraction, $\mathcal{F}(1)|V_{cb}|$ and CLN form factor parameters from the CLN fit are summarised in Table~\ref{systematictable_cln}, while the uncertainties on the BGL fit are given in Table~\ref{systematictable_bgl}. 

We estimate systematic uncertainties by varying each possible uncertainty source such as the PDF shape and signal reconstruction efficiency with the assumption of a Gaussian error, unless otherwise stated. This is done via sets of pseudoexperiments in which each independent systematic uncertainty parameter is randomly varied using a normal distribution. The entire analysis is repeated for each pseudoexperiment and the spread on each measured observable is taken as the systematic error. 

The parameters varied are split into two categories,  those that affect the  shapes and those that affect only the normalization. We start with the former contributions.
\begin{itemize}
\item The tracking efficiency corrections for low momentum tracks vary with track $p_{\rm T}$, as do the relative uncertainties. We conservatively treat the uncertainties in each slow pion $p_{\rm T}$ bin to be fully correlated.
\item The lepton identification efficiencies are varied according to their respective uncertainties, which are dominated by contributions that are correlated across all bins in $p_{\rm lab}$ and $\theta_{\rm lab}$. The electron and muon systematic uncertainties are calculated separately as well as a combined.
\item The results from the background normalization fit are varied within their fitted uncertainties. We take into account finite correlations between the fit results of each component.
\item The uncertainty of the decays ${B} \rightarrow D^{**} \ell^- {\bar \nu}_\ell$ are twofold: the unknown composition of each $D^{**}$ state and the uncertainty in the form-factor parameters used for the MC sample production. The composition uncertainty is estimated based on uncertainties of the branching fractions: $\pm 6\%$ for ${\bar B} \rightarrow D_1 (\rightarrow D^* \pi) \ell \bar{\nu}_\ell$, $\pm 12\%$ for ${\bar B} \rightarrow D_2^* (\rightarrow D^* \pi) \ell \bar{\nu}_\ell$, $\pm 24\%$ for ${\bar B} \rightarrow D'_1 (\rightarrow D^* \pi \pi) \ell \bar{\nu}_\ell$ and $\pm 17\%$ for ${\bar B} \rightarrow D_0^* (\rightarrow D^* \pi) \ell \bar{\nu}_\ell$. If the experimentally-measured branching fractions are not applicable, we vary the branching fractions continuously from $0\%$ to $200\%$ in the MC expectation. We estimate an uncertainty arising from the LLSW model parameters by changing the correction factors within the parameter uncertainties.
\item The relative number of $B^0{\bar B^0}$ meson pairs compared to $B^+B^-$ pairs collected by Belle has a small uncertainty and affects only the relative composition of cross-feed signal events from $B^+$ and $B^0$ decays. The fraction $f_{+-}/f_{00} = \mathcal{B}(\Upsilon(4S) \rightarrow B^+B^-)/\mathcal{B}(\Upsilon(4S) \rightarrow B^0\bar{B^0})$ is varied within its uncertainty~\cite{cite-f+-f00}.
\item Charged hadron identification uncertainties are determined with data using $D^*$ tagged charm decays.
\end{itemize}
The uncertainties that only affect the overall normalization are: the tracking efficiency for high momentum tracks, the branching fraction  ${\cal B}(D^{*+}\to D^0 \pi^+)$ and ${\cal B}(D^{0}\to K^- \pi^+)$, the total number of $\Upsilon(4S)$ events in the sample, and the $B^0$ lifetime. 

\begin{table*}[htb]
 \centering
 \caption{Systematic uncertainty breakdown for $\mathcal{F}(1)|V_{cb}|$, branching fraction  and form factor parameters in the CLN parameterization.}
\label{systematictable_cln}
\setlength{\tabcolsep}{12pt}
 \renewcommand{\arraystretch}{1.6}
\begin{tabularx}{0.92\linewidth}{lrrrcc} 
\hline \hline
Source~~~~&$\rho^{2}$	& $R_{1}(1)$ & $R_{2}(1)$ & $\mathcal{F}(1)|V_{cb}|$ [\%]   &$\mathcal{B}(B^{0} \to D^{*-} \ell^+ \nu_\ell)$ [\%] 	 \\ \hline
Slow pion efficiency                		& 0.005 & 0.002 & 0.001 &0.65 &  1.29 \\ 
Lepton ID combined			        	& 0.001	& 0.006 & 0.004 &0.68 &  1.38 \\ 
$\mathcal B(B \to D^{**} \ell \nu)$ 	&0.002	& 0.001 & 0.002 &0.26 &  0.52 \\ 
$B \to D^{**} \ell \nu$ form factors 	&0.003	& 0.001 & 0.004 &0.11 &  0.22 \\ 
$f_{+-}$/$f_{00}$			        &0.001	& 0.002 & 0.002 &0.52 &  1.06 \\ 
Fake $e/\mu$			            	&0.004	& 0.006 & 0.001 &0.11 &  0.21 \\
Continuum norm.		        &0.002	& 0.002 & 0.001 &0.03 &  0.06 \\
K/$\pi$ ID			    & $<0.001$  & $<0.001$     &$<0.001$      &0.39 &  0.77 \\ 
Fast track efficiency   &- 	    	& -     &-      &0.53 &  1.05 \\
$N\Upsilon(4S)$			&-      	& -     &-      &0.68 &  1.37 \\ 
$B^{0}$ lifetime 		& -	    	& -     &-      &0.13 &  0.26 \\ 
$\mathcal B (D^{*+} \rightarrow D^{0} \pi^{+}_{s})$ &- &- &- & 0.37 &0.74 \\ 
$\mathcal B (D^{0} \rightarrow K \pi )$ &- &- &- & 	0.51 &  1.02 \\ 
\hline
Total systematic error 	&0.008	& 0.009 & 0.007&1.60	& 3.21 \\ 
\hline
\end{tabularx}
\end{table*}
\section{Differential data}\label{differential-data}
In addition to the fit results, we report all necessary data required to perform fits to any choice of form factor parameterization. Specifically we report the background subtracted differential yields ($N_{\rm obs}$) with the statistical error and the signal efficiency ($\epsilon$) in Table~\ref{yield-table}. The systematic uncertainties in each measured bin are in Tables~\ref{sys-table1} - \ref{sys-table4}, the detector response matrices ($R$) in Tables~\ref{response-table1} - \ref{response-table4} for electrons  and \ref{response-table5} - \ref{response-table8} for muons. The statistical uncertainty correlations ($\rho^{\rm stat}$) between measured bins are in Tables~\ref{statcorr-table1-e} - \ref{statcorr-table4-e} for electrons and \ref{statcorr-table1-mu} - \ref{statcorr-table4-mu} for muons. The systematic uncertainty correlations ($\rho^{\rm sys}$) between measured bins are given in Tables~\ref{syscorr-table1} - \ref{syscorr-table4}.  

The correlations between systematic errors in pairs of bins of ($w$, $\cos\theta_{\ell}$, $\cos\theta_{\rm v}$, $\chi$) are determined using a toy MC approach, described in Section~\ref{sec-syst}. The total covariance, for use in the $\chi^2$ minimisation function (Eq.~\ref{chi2}) is defined as
\begin{eqnarray}
{\rm Cov}_{ij} = \rho^{\rm stat}_{ij}\sigma^{\rm stat}_i\sigma^{\rm stat}_j + \rho^{\rm sys}_{ij}\sigma^{\rm sys}_i\sigma^{\rm sys}_j.
\end{eqnarray}
As we provide only the background subtracted differential distributions, the expected yield in Eq. \ref{chi2} becomes
\begin{eqnarray}
N_{i}^{\rm exp.} = \sum_{j=1}^{40} (R_{ij}\epsilon_jN_j^{\rm theory}).
\end{eqnarray}

  The distributions in $w$, $\cos\theta_{\ell}$, $\cos\theta_{\rm v}$ and $\chi$ are divided into 10 bins of equal width where the width of each distribution is equal to 0.05, 0.2, 0.2 and $\frac{2 \pi}{10}$ respectively. The bins are labelled with a common index $i$ where $i$ = 1,...,40. The bins $i$ = 1,...,10 correspond to the 10 bins of $w$ distribution with bin ranging from $w = 1.0$ to $w = 1.50$, $i$ = 11,...,20 correspond to the 10 bins of $\cos\theta_{\ell}$ distribution with bin ranging from $\cos\theta_{\ell}= -1.0$ to $\cos\theta_{\ell} = +1.0$, $i$ = 21,...,30 correspond to the 10 bins of $\cos\theta_{\rm v}$ distribution with bin ranging from $\cos\theta_{\rm v}= -1.0$ to $\cos\theta_{\rm v} = +1.0$ and $i$ = 31,...,40 correspond to the 10 bins of $\chi$ distribution with the bin ranging from $\chi = -\pi$ to $\chi = \pi$. 

  The values of $|V_{cb}|$ and the form factors extracted from fits to these data are found to be compatible with the nominal analysis approach used in this paper. The overall uncertainties may be slightly larger as non-linear correlations of systematic uncertainties are not captured by the covariance matrices.

\section{Results}
The full results for the CLN fit are given below, where the first uncertainty is statistical, and the second systematic:
\begin{eqnarray}
\rho^{2} &=& 1.106 \pm  0.031 \pm 0.007, \\      
R_{1}(1) &=& 1.229  \pm 0.028 \pm 0.009,  \\   
R_{2}(1)  &=& 0.852  \pm 0.021 \pm 0.006, \\
\mathcal{F}(1) \mathcal|V_{cb}|\eta_{\rm EW} \times10^{3} &=& 35.06 \pm 0.15 \pm 0.56,\\
\mathcal{B}(B^{0}\rightarrow D^{*-}\ell^{+}\nu_{\ell}) &=& (4.90 \pm 0.02  \pm 0.16)\%,
\end{eqnarray}
where the first error is statistical and the second error is systematic. The dominant systematic uncertainties are the track reconstruction or the lepton ID uncertainty which are correlated between different bins.  These results are consistent with, and more precise than, those published in Refs.~\cite{cite-dungel, cite-babar1, cite-babar2, cite-babar3}.  We find the value of branching fraction is insensitive to the choice of parameterization. We also present the results from the BGL fit, where the first uncertainty is statistical, and the second systematic.
\begin{eqnarray}
\tilde a_0^f\times10^{3}      & =& -0.506\pm0.004 \pm0.008, \\ 
\tilde a_1^f\times10^{3}      &=&-0.65  \pm 0.17 \pm0.09,   \\
\tilde a_1^{F_1}\times10^{3}  &=&-0.270  \pm 0.064 \pm0.023,    \\
\tilde a_2^{F_1}\times10^{3}  &=&+3.27  \pm 1.25 \pm0.45,    \\
\tilde a_0^g\times10^{3}      &=&-0.929 \pm  0.018 \pm0.013,   \\
\mathcal{F}(1)\mathcal|V_{cb}|\eta_{\rm EW}\times10^{3} &=&  34.93 \pm 0.23 \pm 0.59,\\
\mathcal{B}(B^{0} \to D^{*-} \ell^+ \nu_\ell) &=& (4.90 \pm 0.02 \pm 0.16)\%.
\end{eqnarray}
These results are lower than those based on a preliminary tagged approach by Belle~\cite{cite-belletagged}, as performed in Refs.~\cite{cite-grinstein, cite-gambino}. 
Both sets of fits give acceptable $\chi^2/$ndf: therefore the data do not discriminate between the parameterizations. The result with the BGL paramterisation is consistent with the CLN result but has a larger fit uncertainty.

Taking the value of $\mathcal{F}(1)=0.906\pm 0.013$ from Lattice QCD in Ref.~\cite{cite-LQCD} and $\eta_{\rm EW}=1.0066$ from Ref.~\cite{cite-etaew}, we find the following values for $|V_{cb}|$:
$(38.4 \pm 0.2 \pm 0.6 \pm 0.6)\times10^{-3}$ (CLN+LQCD) and  $(38.3 \pm 0.3 \pm 0.7 \pm 0.6 )\times10^{-3}$ (BGL+LQCD). The errors correspond to the statistical, systematic and lattice QCD uncertainties respectively. The value of $|V_{cb}|$ from the CLN and BGL parameterizations are consistent with the world average and remain to be in tension with inclusive $|V_{cb}|$ value shown in Eq.~\ref{eq:cln_exclusive_val} and Eq.~\ref{eq:inclusive_val} respectively.

We perform a lepton flavor universality (LFU) test by forming a ratio of the branching fractions of modes with electrons and muons. The corresponding value of this ratio is
\begin{eqnarray}
\frac{{\cal B }(B^0 \to D^{*-} e^+ \nu)}{{\cal B }(B^0 \to D^{*-} \mu^+ \nu)} = 1.01 \pm 0.01 \pm 0.03~,
\end{eqnarray}
where the first error is statistical and the second is systematic. The systematic uncertainty is dominated by the electron and muon identification uncertainties, as all others cancel in the ratio. This is the most stringent test of LFU in $B$ decays to date. This result is consistent with unity.

\section{Conclusion}
In this paper we present a new study by the Belle experiment of  $B^{0} \to D^{*-} \ell^+ \nu_\ell$ decay. We present the most precise measurement of $|V_{cb}|$ from exclusive decays, and the first direct measurement using the BGL parameterization. The BGL parameterization gives a value for $|V_{cb}|$ consistent with the CLN parameterization, hence the tension remain with the value from inclusive approach~\cite{cite-belleinclusive1,cite-belleinclusive2,cite-belleinclusive3,cite-hflav}. We also place stringent bounds on lepton flavor universality, as the semi-electronic and semi-muonic branching fractions have been observed to consistent with each other.

\begin{acknowledgements}
We thank the KEKB group for the excellent operation of the
accelerator; the KEK cryogenics group for the efficient
operation of the solenoid; and the KEK computer group, and the Pacific Northwest National
Laboratory (PNNL) Environmental Molecular Sciences Laboratory (EMSL)
computing group for strong computing support; and the National
Institute of Informatics, and Science Information NETwork 5 (SINET5) for
valuable network support.  We acknowledge support from
the Ministry of Education, Culture, Sports, Science, and
Technology (MEXT) of Japan, the Japan Society for the 
Promotion of Science (JSPS), and the Tau-Lepton Physics 
Research Center of Nagoya University; 
the Australian Research Council including grants
DP180102629, 
DP170102389, 
DP170102204, 
DP150103061, 
FT130100303; 
Austrian Science Fund (FWF);
the National Natural Science Foundation of China under Contracts
No.~11435013,  
No.~11475187,  
No.~11521505,  
No.~11575017,  
No.~11675166,  
No.~11705209;  
Key Research Program of Frontier Sciences, Chinese Academy of Sciences (CAS), Grant No.~QYZDJ-SSW-SLH011; 
the  CAS Center for Excellence in Particle Physics (CCEPP); 
the Shanghai Pujiang Program under Grant No.~18PJ1401000;  
the Ministry of Education, Youth and Sports of the Czech
Republic under Contract No.~LTT17020;
the Carl Zeiss Foundation, the Deutsche Forschungsgemeinschaft, the
Excellence Cluster Universe, and the VolkswagenStiftung;
the Department of Science and Technology of India; 
the Istituto Nazionale di Fisica Nucleare of Italy; 
National Research Foundation (NRF) of Korea Grants
No.~2015H1A2A1033649, No.~2016R1D1A1B01010135, No.~2016K1A3A7A09005
603, No.~2016R1D1A1B02012900, No.~2018R1A2B3003 643,
No.~2018R1A6A1A06024970, No.~2018R1D1 A1B07047294; Radiation Science Research Institute, Foreign Large-size Research Facility Application Supporting project, the Global Science Experimental Data Hub Center of the Korea Institute of Science and Technology Information and KREONET/GLORIAD;
the Polish Ministry of Science and Higher Education and 
the National Science Center;
the Grant of the Russian Federation Government, Agreement No.~14.W03.31.0026; 
the Slovenian Research Agency;
Ikerbasque, Basque Foundation for Science, Spain;
the Swiss National Science Foundation; 
the Ministry of Education and the Ministry of Science and Technology of Taiwan;
and the United States Department of Energy and the National Science Foundation.
\end{acknowledgements}

\begin{appendix}
\begin{table*}[htb]
 \centering
 \caption{Systematic uncertainty breakdown for $\mathcal{F}(1)|V_{cb}|$, branching fraction  and form factor parameters in the BGL parameterization.}
\label{systematictable_bgl}
\setlength{\tabcolsep}{6pt}
 \renewcommand{\arraystretch}{1.6}
\begin{tabularx}{1.0\linewidth}{lccccccc} 
\hline \hline
Source~~~~&$\tilde{a}_{0}^{f}$ [\%] & $\tilde{a}_{1}^{f}$ [\%] & $\tilde{a}_{1}^{F1}$ [\%] & $\tilde{a}_{2}^{F1}$ [\%]& $\tilde{a}_{0}^{g}$ [\%]& $\eta_{EW}\mathcal{F}(1)|V_{cb}|$  [\%] &$\mathcal{B}(B^{0} \to D^{*-} \ell^+ \nu_{\ell})$ [\%]  	 \\ \hline
Slow pion efficiency &0.79&9.59&5.61&4.46&0.18&0.79&1.57\\
Lepton ID combined  &0.67&5.45&1.35&0.73&0.38&0.67&1.33\\
$\mathcal B(B \to D^{**} \ell \nu)$ &0.05&5.02&4.34&9.31&0.37&0.05 &0.10\\
$B \to D^{**} \ell \nu$ form factors &0.08&2.08&3.56&6.78&0.12&0.08&0.16\\
$f_{+-}$/$f_{00}$ &0.56&0.46&0.50&0.48&0.56&0.56&1.05\\
Fake $e/\mu$ &0.07&6.43&3.03&5.92&0.14&0.07&0.11\\
K/$\pi$ ID & 0.39 & 0.39 &0.39&0.39  & 0.39 &0.39  &0.77\\
Fast track efficiency   &0.53 	& 0.53     &0.53   &0.53 &0.53 &0.53 &  1.05 \\
$N(\Upsilon(4S))$	& 0.69     	& 0.69     &0.69    &0.69 &0.69  &0.69 &  1.37 \\ 
$B^{0}$ lifetime 	& 0.13	& 0.13     &0.13     &0.13 &0.13&0.13 &  0.26 \\ 
$\mathcal B (D^{*+} \rightarrow D^{0} \pi^{+}_{s})$ &0.37 &0.37 &0.37 &0.37 &0.37& 0.37 &0.74 \\ 
$\mathcal B (D^{0} \rightarrow K \pi )$ &0.51 &0.51 &0.51 &0.51 &0.51& 	0.51 &  1.02 \\ 
\hline
Total systematic error &1.65&13.93&8.69&13.77&1.40&1.65&3.26 \\
\hline
\end{tabularx}
\end{table*}
\onecolumngrid
\begin{table*}[htb]
 \centering
\caption{Background subtracted signal yield and selection efficiency in the 40 bins defined in Section~\ref{differential-data}. The left (right) part of the table is for the electron (muon) mode. Only statistical uncertainty is quoted.}
\label{yield-table}
\setlength{\tabcolsep}{12pt}
\renewcommand{\arraystretch}{1.5}
\begin{tabularx}{0.70\linewidth}{c|rr|rr}
\hline \hline
Bin   & Yield  			& Efficiency ($\%$)  	 & Yield	  	& Efficiency ($\%$)  \\ \hline
1	&1421 $\pm$ 41 	&2.72 $\pm$ 0.02	&1494 $\pm$ 43	& 2.68 $\pm$ 0.02	 \\
2	&5319 $\pm$ 85 	&5.72 $\pm$ 0.02	&5062 $\pm$ 89	& 5.66 $\pm$ 0.02	 \\
3	&8563 $\pm$ 113 	&7.70 $\pm$ 0.03	&8385 $\pm$ 120	& 7.66 $\pm$ 0.03	 \\
4	&10685 $\pm$ 129 	&9.10 $\pm$ 0.03	&10734 $\pm$ 142	& 9.05 $\pm$ 0.03	 \\
5	&11971 $\pm$ 156 	&10.03 $\pm$ 0.03	&11961 $\pm$ 159	& 9.91 $\pm$ 0.03	 \\
6	&12275 $\pm$ 167 	&10.61 $\pm$ 0.03	&12090 $\pm$ 167	& 10.43 $\pm$ 0.03	 \\
7	&11888 $\pm$ 166 	&10.74 $\pm$ 0.03	&11803 $\pm$ 168	& 10.60 $\pm$ 0.03	 \\
\hline
\end{tabularx}
\end{table*}

\begin{table*}[htb]
\caption{Systematic uncertainty ($\%$) in each bin of the observable $w$. The bins are defined in Section~\ref{differential-data}.}
\label{sys-table1}
\setlength{\tabcolsep}{6pt}
\renewcommand{\arraystretch}{1.5}
\begin{tabularx}{0.8\linewidth}{lYYYYYYYYYY}
\hline \hline
Source  &1& 2 &  3 & 4&  5&6 &7&8&9&10\\ \hline
$\mathcal{B}(D^{0}\rightarrow K \pi$) &1.02 &	1.02 &	1.02 &	1.01 &	1.01 &	1.02 &	1.02 &	1.02 &	1.02 &	1.02 \\
$\mathcal{B}(D^{*+}\rightarrow D^{0} \pi^{+}_{s}$) &0.74 &	0.74 &	0.74 &	0.74 &	0.74 &	0.74 &	0.74 &	0.74 &	0.74 &	0.71 \\
Lepton ID(e) 		&1.38 &	1.48 &	1.58 &	1.57 &	1.80 &	1.89 &	1.90 &	2.02 &	2.04 &	2.05 \\ 
Lepton ID($\mu$)	&2.23 &	2.12 &	2.05 &	2.01 &	2.04 &	2.05 &	2.04 &	2.03 &	1.93 &	1.93 \\ 
Lepton ID	&1.18 &1.21 &1.25 &1.24 &1.35 &1.39 &1.39 &1.43 &1.40 &1.41 \\
Slow track efficiency&5.77 &	3.01 &	2.14 &	1.75 &	1.53 &	1.38 &	1.33 &	1.26 &	1.12 &	0.84 \\
$e/\mu$ fake rate  &0.03 &0.01 &0.04 &0.06 &0.12 &	0.12 &	0.13 &	0.17 &	0.27 &	0.17 \\
$D^{**}$ branching fraction  &0.44 &	0.15 &	0.01 &	0.41 &	0.06 &	0.04 &	0.08 &	0.60 &	0.35 &	0.22 \\
$D^{**}$ shape &0.02 &	0.11 &	0.14 &	0.01 &	0.16 &	0.30 &	0.22 &	0.08 &	0.35 &	0.92 \\
$f_{+-}/f_{00}$ &1.05 &	1.07 &	1.10 &	1.10 &	1.09 &	1.08 &	1.11 &	1.08 &	1.05 &	1.08 \\
Norm. continuum &0.06 &	0.06 &	0.06 &	0.06 &	0.06 &	0.06 &	0.06 &	0.06 &	0.06 &	0.06 \\
Fast track efficiency &1.05 &	1.05 &	1.05 &	1.05 &	1.05 &	1.05 &	1.05 &	1.05 &	1.05 &	1.05 \\
$N(\Upsilon(4S))$  &1.37 &	1.37 &	1.37 &	1.37 &	1.37 &	1.37 &	1.37 &	1.37 &	1.37 &	1.37 \\
$B^{0}$ lifetime &0.26 &	0.26 &	0.26 &	0.26 &	0.26 &	0.26 &	0.26 &	0.26 &	0.26 &	0.26 \\
$K/\pi$ ID &0.77 &	0.77 &	0.77 &	0.77 &	0.77 &	0.77 &	0.77 &	0.77 &	0.77 &	0.77 \\
\hline
Total& 6.42 &	 4.12 &	 3.55 &	 3.35 &	 3.26 &	 3.22 &	 3.20 &	 3.23 &	 3.14 &	 3.16 \\
\hline
\end{tabularx}
\end{table*}

\begin{table*}[htb]
\caption{Systematic uncertainty ($\%$) in each bin of the observable $\cos\theta_{\ell}$. The bins are defined in Section~\ref{differential-data}.}
\label{sys-table2}
\setlength{\tabcolsep}{6pt}
\renewcommand{\arraystretch}{1.5}
\begin{tabularx}{0.8\linewidth}{lYYYYYYYYYY}
\hline \hline
Source    &11& 12 & 13 & 14&  15&16 &17&18&19&20	\\  \hline
$\mathcal{B}(D^{0}\rightarrow K \pi$) &1.02 &	1.02 &	1.02 &	1.03 &	1.02 &	1.01 &	1.02 &	1.02 &	1.02 &	1.02 \\
$\mathcal{B}(D^{*+}\rightarrow D^{0} \pi^{+}_{s}$) &0.74 &	0.74 &	0.74 &	0.74 &	0.74 &	0.74 &	0.74 &	0.74 &	0.74 &	0.74 \\
Lepton ID(e)					&4.26 &	4.07 &	3.54 &	2.66 &	1.94 &	1.41 &	1.43 &	1.40 &	1.46 &	1.52 \\ 
Lepton ID($\mu$)				&2.52 &	2.67 &	2.60 &	2.18 &	2.04 &	2.05 &	1.93 &	1.95 &	1.94 &	1.74 \\
Lepton ID&2.17 &2.23 &2.09 &1.68 &1.41 &1.16 &1.15 &1.14 &1.17 &1.14  \\
Slow track efficiency&2.83 &	1.95 &	1.49 &	1.28 &	1.27 &	1.30 &	1.33 &	1.38 &	1.45 &	1.52 \\
$e/\mu$ fake rate  &0.30 &	0.13 &	0.05 &	0.10 &	0.12 &	0.14 &	0.15 &	0.15 &	0.11 &	0.13 \\
$D^{**}$ branching fraction  &0.15 &	0.13 &	0.11 &	0.15 &	0.10 &	0.05 &	0.06 &	0.08 &	0.05 &	0.08 \\
$D^{**}$ shape &0.10 &	0.10 &	0.15 &	0.11 &	0.10 &	0.14 &	0.06 &	0.08 &	0.02 &	0.07 \\
$f_{+-}/f_{00}$ &1.08 &	1.08 &	1.08 &	1.07 &	1.08 &	1.07 &	1.09 &	1.09 &	1.08 &	1.05 \\
Norm. continuum &0.06 &	0.06 &	0.06 &	0.06 &	0.06 &	0.06 &	0.06 &	0.06 &	0.06 &	0.06 \\
Fast track efficiency &1.05 &	1.05 &	1.05 &	1.05 &	1.05 &	1.05 &	1.05 &	1.05 &	1.05 &	1.05 \\
$N(\Upsilon(4S))$  &1.37 &	1.37 &	1.37 &	1.37 &	1.37 &	1.37 &	1.37 &	1.37 &	1.37 &	1.37 \\
$B^{0}$ lifetime &0.26 &	0.26 &	0.26 &	0.26 &	0.26 &	0.26 &	0.26 &	0.26 &	0.26 &	0.26 \\
$K/\pi$ ID &0.77 &	0.77 &	0.77 &	0.77 &	0.77 &	0.77 &	0.77 &	0.77 &	0.77 &	0.77 \\
\hline
Total & 4.39 &	 3.90 &	 3.61 &	 3.30 &	 3.17 &	 3.07 &	 3.09 &	 3.10 &	 3.14 &	 3.16 \\
\hline
\end{tabularx}
\end{table*}

\begin{table*}[htb]
\caption{Systematic uncertainty ($\%$) in each bin of the observable $\cos\theta_{\rm v}$. The bins are defined in Section~\ref{differential-data}.}
\label{sys-table3}
\setlength{\tabcolsep}{6pt}
\renewcommand{\arraystretch}{1.5}
\begin{tabularx}{0.8\linewidth}{lYYYYYYYYYY}
\hline \hline
 Source   &21&22 &  23 & 24&  25&26 &27&28&29&30\\  \hline
$\mathcal{B}(D^{0}\rightarrow K \pi$) &1.02 &	1.02 &	1.02 &	1.02 &	1.02 &	1.02 &	1.02 &	1.02 &	1.02 &	1.02 \\
$\mathcal{B}(D^{*+}\rightarrow D^{0} \pi^{+}_{s}$) &0.74 &	0.74 &	0.74 &	0.74 &	0.74 &	0.74 &	0.74 &	0.74 &	0.74 &	0.74 \\
Lepton ID(e)					&1.95 &	1.91 &	1.83 &	1.72 &	1.62 &	1.65 &	1.72 &	1.83 &	1.90 &	1.94 \\ 
Lepton ID($\mu$)				&2.15 &	2.13 &	2.09 &	2.04 &	2.05 &	1.90 &	1.96 &	1.93 &	1.90 &	1.86 \\ 
Lepton ID &1.44 &1.42 &1.38 &1.31 &1.27 &1.24 &1.29 &1.33 &1.34 &1.34  \\
Slow track efficiency&1.02 &	1.14 &	1.28 &	1.39 &	1.52 &	1.68 &	1.84 &	1.99 &	2.18 &	2.63 \\
$e/\mu$ fake rate  &0.04 &	0.06 &	0.11 &	0.16 &	0.22 &	0.19 &	0.18 &	0.17 &	0.21 &	0.04 \\
$D^{**}$ branching fraction  &0.08 &	0.01 &	0.17 &	0.32 &	0.27 &	0.19 &	0.07 &	0.14 &	0.18 &	0.33 \\
$D^{**}$ shape &0.07 &	0.03 &	0.02 &	0.05 &	0.16 &	0.12 &	0.05 &	0.19 &	0.00 &	0.08 \\
$f_{+-}/f_{00}$ &1.08 &	1.08 &	1.07 &	1.09 &	1.07 &	1.09 &	1.10 &	1.08 &	1.10 &	1.08 \\
Norm. continuum &0.06 &	0.06 &	0.06 &	0.06 &	0.06 &	0.06 &	0.06 &	0.06 &	0.06 &	0.06 \\
Fast track efficiency &1.05 &	1.05 &	1.05 &	1.05 &	1.05 &	1.05 &	1.05 &	1.05 &	1.05 &	1.05 \\
$N(\Upsilon(4S))$ &1.37 &	1.37 &	1.37 &	1.37 &	1.37 &	1.37 &	1.37 &	1.37 &	1.37 &	1.37 \\
$B^{0}$ lifetime &0.26 &	0.26 &	0.26 &	0.26 &	0.26 &	0.26 &	0.26 &	0.26 &	0.26 &	0.26 \\
$K/\pi$ ID &0.77 &	0.77 &	0.77 &	0.77 &	0.77 &	0.77 &	0.77 &	0.77 &	0.77 &	0.77 \\
\hline
 Total &3.09 &	 3.12 &	 3.16 &	 3.19 &	 3.23 &	 3.30 &	 3.39 &	 3.49 &	 3.62 &	 3.90 \\
\hline
\end{tabularx}
\end{table*}

\begin{table*}[htb]
\caption{Systematic uncertainty ($\%$) in each bin of the observable $\chi$. The bins are defined in Section~\ref{differential-data}.}
\label{sys-table4}
\setlength{\tabcolsep}{6pt}
\renewcommand{\arraystretch}{1.5}
\begin{tabularx}{0.8\linewidth}{lYYYYYYYYYY}
\hline \hline
  Source  	&31&32 &  33 & 34&  35&36 &37&38&39&40 \\  \hline
$\mathcal{B}(D^{0}\rightarrow K \pi$) &1.02 &	1.02 &	1.02 &	1.02 &	1.02 &	1.02 &	1.02 &	1.02 &	1.02 &	1.02 \\
$\mathcal{B}(D^{*+}\rightarrow D^{0} \pi^{+}_{s}$)&0.74 &	0.74 &	0.74 &	0.74 &	0.74 &	0.74 &	0.74 &	0.74 &	0.74 &	0.74 \\
 Lepton ID(e)					&1.81 &	1.77 &	1.83 &	1.80 &	1.82 &	1.84 &	1.85 &	1.86 &	1.83 &	1.82 \\
Lepton ID($\mu$)				&1.89 &	1.97 &	2.08 &	2.06 &	2.09 &	2.08 &	2.12 &	1.99 &	1.95 &	1.87 \\
Lepton ID &1.31 &1.32 &1.38 &1.36 &1.37 &1.38 &1.39 &1.36 &1.34 &1.30 \\
Slow track efficiency &1.47 &	1.45 &	1.40 &	1.33 &	1.28 &	1.31 &	1.36 &	1.45 &	1.47 &	1.46 \\
$e/\mu$ fake rate &0.15 &	0.10 &	0.09 &	0.12 &	0.16 &	0.11 &	0.09 &	0.13 &	0.13 &	0.16 \\
$D^{**}$ branching fraction &0.15 &	0.10 &	0.01 &	0.03 &	0.06 &	0.02 &	0.03 &	0.13 &	0.02 &	0.16 \\
$D^{**}$ shape  &0.01 &	0.07 &	0.01 &	0.09 &	0.08 &	0.07 &	0.01 &	0.13 &	0.13 &	0.00 \\
$f_{+-}/f_{00}$ &11.08 &	1.07 &	1.09 &	1.06 &	1.08 &	1.07 &	1.09 &	1.08 &	1.06 &	1.10 \\
Norm. continuum &0.06 &	0.06 &	0.06 &	0.06 &	0.06 &	0.06 &	0.06 &	0.06 &	0.06 &	0.06 \\
Fast track efficiency &1.05 &	1.05 &	1.05 &	1.05 &	1.05 &	1.05 &	1.05 &	1.05 &	1.05 &	1.05 \\
$N(\Upsilon(4S))$  &1.37 &	1.37 &	1.37 &	1.37 &	1.37 &	1.37 &	1.37 &	1.37 &	1.37 &	1.37 \\
$B^{0}$ lifetime &0.26 &	0.26 &	0.26 &	0.26 &	0.26 &	0.26 &	0.26 &	0.26 &	0.26 &	0.26 \\
$K/\pi$ ID &0.77 &	0.77 &	0.77 &	0.77 &	0.77 &	0.77 &	0.77 &	0.77 &	0.77 &	0.77 \\
\hline
Total& 3.21 &	 3.20 &	 3.20 &	 3.16 &	 3.16 &	 3.16 &	 3.20 &	 3.22 &	 3.22 &	 3.21 \\
\hline
\end{tabularx}
\end{table*}

\begin{table*}[htb]
\caption{Response matrix $R$ for observable $w$ for the electron mode. The bins are defined in Section~\ref{differential-data}.} 
\label{response-table1}
\setlength{\tabcolsep}{6pt}
\renewcommand{\arraystretch}{1.5}
\begin{tabularx}{0.7\linewidth}{c|cccccccccc}
\hline \hline
Bin &1      &2      &3      &4      &5      &6      &7      &8      &9      &10\\
\hline
1  &0.803  &0.053  &0.000  &0.000  &0.000  &0.000  &0.000  &0.000  &0.000  &0.000 \\
2  &0.197  &0.778  &0.098  &0.000  &0.000  &0.000  &0.000  &0.000  &0.000  &0.000 \\
3  &0.000  &0.168  &0.717  &0.126  &0.002  &0.000  &0.000  &0.000  &0.000  &0.000 \\
4  &0.000  &0.001  &0.182  &0.667  &0.149  &0.006  &0.000  &0.000  &0.000  &0.000 \\
5  &0.000  &0.000  &0.004  &0.199  &0.626  &0.167  &0.011  &0.000  &0.000  &0.000 \\
6  &0.000  &0.000  &0.000  &0.009  &0.207  &0.592  &0.177  &0.015  &0.000  &0.000 \\
7  &0.000  &0.000  &0.000  &0.000  &0.016  &0.215  &0.575  &0.183  &0.018  &0.000 \\
8  &0.000  &0.000  &0.000  &0.000  &0.000  &0.021  &0.213  &0.567  &0.186  &0.017 \\
9  &0.000  &0.000  &0.000  &0.000  &0.000  &0.000  &0.024  &0.214  &0.598  &0.186 \\
10 &0.000  &0.000  &0.000  &0.000  &0.000  &0.000  &0.000  &0.022  &0.198  &0.797 \\
\hline
\end{tabularx}
\end{table*}

\begin{table*}[htb]
\caption{Response matrix $R$ for observable $\cos\theta_{\ell}$ for the electron mode. The bins are defined in Section~\ref{differential-data}.} 
\label{response-table2}
\setlength{\tabcolsep}{6pt}
\renewcommand{\arraystretch}{1.5}
\begin{tabularx}{0.7\linewidth}{c|cccccccccc}
\hline \hline
Bin &1      &2      &3      &4      &5      &6      &7      &8      &9      &10\\
\hline
1 &0.961  &0.024  &0.000  &0.000  &0.000  &0.000  &0.000  &0.000  &0.000  &0.000 \\
2 &0.038  &0.952  &0.027  &0.000  &0.000  &0.000  &0.000  &0.000  &0.000  &0.000 \\
3 &0.000  &0.021  &0.948  &0.041  &0.001  &0.000  &0.000  &0.000  &0.000  &0.000 \\
4 &0.000  &0.001  &0.023  &0.918  &0.067  &0.003  &0.001  &0.001  &0.001  &0.000 \\
5 &0.000  &0.001  &0.001  &0.040  &0.871  &0.097  &0.005  &0.001  &0.001  &0.000 \\
6 &0.000  &0.000  &0.000  &0.001  &0.060  &0.817  &0.129  &0.006  &0.001  &0.000 \\
7 &0.000  &0.000  &0.000  &0.000  &0.001  &0.082  &0.758  &0.164  &0.007  &0.001 \\
8 &0.000  &0.000  &0.000  &0.000  &0.000  &0.001  &0.106  &0.698  &0.196  &0.008 \\
9 &0.000  &0.000  &0.000  &0.000  &0.000  &0.000  &0.001  &0.128  &0.657  &0.212 \\
10 &0.000  &0.000  &0.000  &0.000  &0.000  &0.000  &0.000  &0.002  &0.137  &0.777 \\
\hline
\end{tabularx}
\end{table*}

\begin{table*}[htb]
\caption{Response matrix $R$ for observable $\cos\theta_{\rm v}$ for the electron mode. The bins are defined in Section~\ref{differential-data}.}
\label{response-table3}
\setlength{\tabcolsep}{6pt}
\renewcommand{\arraystretch}{1.5}
\begin{tabularx}{0.7\linewidth}{c|cccccccccc}
\hline \hline
Bin &1      &2      &3      &4      &5      &6      &7      &8      &9      &10\\
\hline
1 &0.918  &0.077  &0.000  &0.000  &0.000  &0.000  &0.000  &0.000  &0.000  &0.000 \\
2 &0.082  &0.806  &0.095  &0.001  &0.000  &0.000  &0.000  &0.000  &0.000  &0.000 \\
3 &0.000  &0.115  &0.761  &0.101  &0.002  &0.000  &0.000  &0.000  &0.000  &0.000 \\
4 &0.000  &0.001  &0.141  &0.735  &0.105  &0.002  &0.000  &0.000  &0.000  &0.000 \\
5 &0.000  &0.000  &0.002  &0.160  &0.719  &0.100  &0.001  &0.000  &0.000  &0.000 \\
6 &0.000  &0.000  &0.000  &0.003  &0.170  &0.722  &0.093  &0.001  &0.000  &0.000 \\
7 &0.000  &0.000  &0.000  &0.000  &0.003  &0.173  &0.738  &0.080  &0.001  &0.000 \\
8 &0.000  &0.000  &0.000  &0.000  &0.000  &0.002  &0.166  &0.771  &0.072  &0.000 \\
9 &0.000  &0.000  &0.000  &0.000  &0.000  &0.000  &0.001  &0.147  &0.819  &0.064 \\
10 &0.000  &0.000  &0.000  &0.000  &0.000  &0.000  &0.000  &0.001  &0.108  &0.936 \\
\hline
\end{tabularx}
\end{table*}

\begin{table*}[htb]
\caption{Response matrix $R$ for observable $\chi$ for the electron mode. The bins are defined in Section~\ref{differential-data}.}
\label{response-table4}
\setlength{\tabcolsep}{6pt}
\renewcommand{\arraystretch}{1.5}
\begin{tabularx}{0.7\linewidth}{c|cccccccccc}
\hline \hline
Bin &1      &2      &3      &4      &5      &6      &7      &8      &9      &10\\
\hline
 1 &0.659  &0.129  &0.011  &0.003  &0.002  &0.002  &0.002  &0.004  &0.013  &0.144 \\
 2&0.151  &0.691  &0.132  &0.012  &0.004  &0.002  &0.002  &0.002  &0.004  &0.016 \\
 3&0.015  &0.141  &0.697  &0.147  &0.016  &0.005  &0.002  &0.002  &0.002  &0.005 \\
 4&0.005  &0.012  &0.134  &0.671  &0.162  &0.018  &0.005  &0.002  &0.002  &0.002 \\
 5&0.002  &0.004  &0.013  &0.140  &0.634  &0.155  &0.016  &0.004  &0.002  &0.002 \\
 6&0.002  &0.002  &0.004  &0.015  &0.155  &0.633  &0.141  &0.013  &0.004  &0.003 \\
 7&0.002  &0.002  &0.002  &0.004  &0.018  &0.163  &0.670  &0.136  &0.012  &0.004 \\
 8&0.005  &0.002  &0.002  &0.002  &0.005  &0.015  &0.147  &0.695  &0.140  &0.015 \\
 9&0.016  &0.004  &0.002  &0.002  &0.002  &0.004  &0.013  &0.132  &0.691  &0.150 \\
 10&0.142  &0.013  &0.003  &0.002  &0.002  &0.002  &0.003  &0.012  &0.130  &0.659 \\
 \hline
\end{tabularx}
\end{table*}

\begin{table*}[htb]
\caption{Response matrix $R$ for observable $w$ for the muon mode. The bins are defined in Section~\ref{differential-data}.}
\label{response-table5}
\setlength{\tabcolsep}{6pt}
\renewcommand{\arraystretch}{1.5}
\begin{tabularx}{0.7\linewidth}{c|cccccccccc}
\hline \hline
Bin &1      &2      &3      &4      &5      &6      &7      &8      &9      &10\\
\hline
1 &0.812  &0.051  &0.000  &0.000  &0.000  &0.000  &0.000  &0.000  &0.000  &0.000 \\
2 &0.188  &0.784  &0.096  &0.000  &0.000  &0.000  &0.000  &0.000  &0.000  &0.000 \\
3 &0.000  &0.164  &0.728  &0.126  &0.002  &0.000  &0.000  &0.000  &0.000  &0.000 \\
4 &0.000  &0.001  &0.172  &0.676  &0.149  &0.006  &0.000  &0.000  &0.000  &0.000 \\
5 &0.000  &0.000  &0.004  &0.190  &0.631  &0.165  &0.010  &0.000  &0.000  &0.000 \\
6 &0.000  &0.000  &0.000  &0.008  &0.203  &0.600  &0.181  &0.016  &0.000  &0.000 \\
7 &0.000  &0.000  &0.000  &0.000  &0.014  &0.209  &0.578  &0.187  &0.019  &0.000 \\
8 &0.000  &0.000  &0.000  &0.000  &0.000  &0.020  &0.209  &0.573  &0.195  &0.017 \\
9 &0.000  &0.000  &0.000  &0.000  &0.000  &0.000  &0.022  &0.205  &0.600  &0.195 \\
10 &0.000  &0.000  &0.000  &0.000  &0.000  &0.000  &0.000  &0.019  &0.186  &0.788 \\
\hline
\end{tabularx}
\end{table*}

\begin{table*}[htb]
\caption{Response matrix $R$ for observable $\cos\theta_{\ell}$ for the muon mode. The bins are defined in Section~\ref{differential-data}.}
\label{response-table6}
\setlength{\tabcolsep}{6pt}
\renewcommand{\arraystretch}{1.5}
\begin{tabularx}{0.7\linewidth}{c|cccccccccc}
\hline \hline
Bin &1      &2      &3      &4      &5      &6      &7      &8      &9      &10\\
\hline
1 &0.959  &0.022  &0.000  &0.000  &0.000  &0.000  &0.000  &0.000  &0.000  &0.000 \\
2 &0.039  &0.955  &0.012  &0.000  &0.000  &0.000  &0.000  &0.000  &0.000  &0.000 \\
3 &0.000  &0.021  &0.960  &0.022  &0.001  &0.000  &0.000  &0.000  &0.000  &0.000 \\
4 &0.001  &0.001  &0.026  &0.931  &0.043  &0.001  &0.000  &0.000  &0.000  &0.000 \\
5 &0.000  &0.000  &0.001  &0.047  &0.889  &0.070  &0.002  &0.000  &0.000  &0.000 \\
6 &0.000  &0.001  &0.000  &0.000  &0.067  &0.837  &0.103  &0.002  &0.001  &0.000 \\
7 &0.000  &0.000  &0.000  &0.000  &0.000  &0.091  &0.778  &0.138  &0.003  &0.000 \\
8 &0.000  &0.000  &0.000  &0.000  &0.000  &0.000  &0.117  &0.715  &0.174  &0.004 \\
9 &0.000  &0.000  &0.000  &0.000  &0.000  &0.000  &0.001  &0.142  &0.672  &0.193 \\
10 &0.000  &0.000  &0.000  &0.000  &0.000  &0.000  &0.000  &0.002  &0.151  &0.803 \\
\hline
\end{tabularx}
\end{table*}

\begin{table*}[htb]
\caption{Response matrix $R$ for observable $\cos\theta_{\rm v}$ for the muon mode. The bins are defined in Section~\ref{differential-data}.}
\label{response-table7}
\setlength{\tabcolsep}{6pt}
\renewcommand{\arraystretch}{1.5}
\begin{tabularx}{0.7\linewidth}{c|cccccccccc}
\hline \hline
Bin &1      &2      &3      &4      &5      &6      &7      &8      &9      &10\\
\hline
1 &0.918  &0.077  &0.000  &0.000  &0.000  &0.000  &0.000  &0.000  &0.000  &0.000 \\
2 &0.082  &0.805  &0.091  &0.001  &0.000  &0.000  &0.000  &0.000  &0.000  &0.000 \\
3 &0.000  &0.117  &0.763  &0.101  &0.002  &0.000  &0.000  &0.000  &0.000  &0.000 \\
4 &0.000  &0.001  &0.142  &0.735  &0.103  &0.002  &0.000  &0.000  &0.000  &0.000 \\
5 &0.000  &0.000  &0.003  &0.159  &0.723  &0.098  &0.001  &0.000  &0.000  &0.000 \\
6 &0.000  &0.000  &0.000  &0.004  &0.169  &0.726  &0.091  &0.001  &0.000  &0.000 \\
7 &0.000  &0.000  &0.000  &0.000  &0.004  &0.172  &0.745  &0.082  &0.001  &0.000 \\
8 &0.000  &0.000  &0.000  &0.000  &0.000  &0.002  &0.161  &0.771  &0.074  &0.000 \\
9 &0.000  &0.000  &0.000  &0.000  &0.000  &0.000  &0.001  &0.145  &0.817  &0.066 \\
10 &0.000  &0.000  &0.000  &0.000  &0.000  &0.000  &0.000  &0.000  &0.107  &0.934 \\
\hline
\end{tabularx}
\end{table*}

\begin{table*}[htb]
\caption{Response matrix $R$ for observable $\chi$ for the muon mode. The bins are defined in Section~\ref{differential-data}.}
\label{response-table8}
\setlength{\tabcolsep}{6pt}
\renewcommand{\arraystretch}{1.5}
\begin{tabularx}{0.7\linewidth}{c|cccccccccc}
\hline \hline
Bin &1      &2      &3      &4      &5      &6      &7      &8      &9      &10\\
\hline
1 &0.653 &0.129 &0.012 &0.004 &0.003 &0.002 &0.002 &0.004 &0.014 &0.144 \\
2 &0.152 &0.686 &0.130 &0.013 &0.004 &0.003 &0.002 &0.002 &0.005 &0.017 \\
3 &0.016 &0.143 &0.693 &0.147 &0.016 &0.006 &0.003 &0.002 &0.003 &0.005 \\
4 &0.005 &0.013 &0.138 &0.667 &0.160 &0.018 &0.005 &0.002 &0.002 &0.003 \\
5 &0.003 &0.004 &0.013 &0.142 &0.630 &0.156 &0.015 &0.004 &0.002 &0.002 \\
6 &0.002 &0.002 &0.004 &0.015 &0.158 &0.629 &0.142 &0.013 &0.004 &0.003 \\
7 &0.003 &0.002 &0.002 &0.005 &0.018 &0.164 &0.667 &0.138 &0.013 &0.005 \\
8 &0.005 &0.003 &0.002 &0.003 &0.006 &0.016 &0.148 &0.692 &0.141 &0.016 \\
9 &0.017 &0.004 &0.002 &0.002 &0.003 &0.005 &0.013 &0.131 &0.686 &0.152 \\
10 &0.144 &0.014 &0.004 &0.002 &0.002 &0.002 &0.004 &0.012 &0.129 &0.654 \\
\hline
\end{tabularx}
\end{table*}

\begin{sidewaystable*}[htb]
\tiny
\caption{Statistical uncertainty correlation matrix of $(w, \cos\theta_{\ell})$ versus $(w, \cos\theta_{\ell})$ for electron mode. The bins are defined in Section~\ref{differential-data}.}
\label{statcorr-table1-e}
\centering
\setlength{\tabcolsep}{4pt}
\renewcommand{\arraystretch}{1.5}
 \begin{adjustbox}{width=\textwidth}
\begin{tabularx}{0.8\linewidth}{r|rrrrrrrrrrrrrrrrrrrr}
\hline \hline
Bin &1      &2      &3      &4      &5      &6      &7      &8      &9      &10 &11      &12      &13      &14      &15      &16      &17      &18      &19      &20 \\
\hline
1  &1.000  &0.000  &0.000  &0.000  &0.000  &0.000  &0.000  &0.000  &0.000  &0.000  &0.016  &0.012  &0.007  &0.002  &-0.003  &-0.002  &-0.004  &-0.006  &-0.003  &0.007 \\
2  &0.000  &1.000  &0.000  &0.000  &0.000  &0.000  &0.000  &0.000  &0.000  &0.000  &0.032  &0.015  &0.005  &-0.005  &-0.010  &-0.009  &-0.005  &-0.002  &0.003  &0.019 \\
3  &0.000  &0.000  &1.000  &0.000  &0.000  &0.000  &0.000  &0.000  &0.000  &0.000  &0.031  &0.016  &0.007  &-0.005  &-0.010  &-0.010  &-0.007  &-0.003  &0.004  &0.028 \\
4  &0.000  &0.000  &0.000  &1.000  &0.000  &0.000  &0.000  &0.000  &0.000  &0.000  &0.022  &0.011  &0.008  &-0.005  &-0.011  &-0.008  &-0.005  &-0.001  &0.008  &0.030 \\
5  &0.000  &0.000  &0.000  &0.000  &1.000  &0.000  &0.000  &0.000  &0.000  &0.000  &0.015  &0.008  &0.001  &-0.005  &-0.008  &-0.006  &-0.003  &0.002  &0.009  &0.024 \\
6  &0.000  &0.000  &0.000  &0.000  &0.000  &1.000  &0.000  &0.000  &0.000  &0.000  &-0.001  &0.005  &0.002  &-0.001  &-0.002  &-0.000  &0.000  &0.002  &0.003  &0.010 \\
7  &0.000  &0.000  &0.000  &0.000  &0.000  &0.000  &1.000  &0.000  &0.000  &0.000  &-0.016  &0.001  &0.001  &0.001  &0.001  &0.003  &0.002  &0.003  &0.004  &0.001 \\
\hline
\end{tabularx}
 \end{adjustbox}
\end{sidewaystable*}

\begin{sidewaystable*}[htb]
\tiny
\caption{Statistical uncertainty correlation matrix of $(\cos\theta_{\rm v}, \chi)$ versus $(w,\cos\theta_{\ell})$ for the electron mode. The bins are defined in Section~\ref{differential-data}.}
\label{statcorr-table2-e}
\centering
\setlength{\tabcolsep}{4pt}
\renewcommand{\arraystretch}{1.5}
 \begin{adjustbox}{width=\textwidth}
\begin{tabularx}{0.8\linewidth}{c|rrrrrrrrrrrrrrrrrrrr}
\hline \hline
Bin &21      &22      &23      &24      &25      &26      &27      &28      &29      &30 &31      &32      &33      &34      &35      &36      &37      &38      &39      &40 \\
\hline
1  &0.022  &0.009  &0.002  &-0.001  &-0.003  &-0.004  &-0.003  &0.001  &0.004  &0.004  &-0.003  &0.005  &0.012  &0.019  &0.017  &0.015  &0.007  &-0.004  &-0.026  &-0.057 \\
2  &0.019  &0.021  &0.013  &0.006  &0.001  &-0.005  &-0.003  &-0.006  &-0.006  &-0.004  &0.000  &0.003  &0.006  &0.007  &0.006  &0.004  &0.002  &-0.001  &-0.007  &-0.019 \\
3  &0.017  &0.027  &0.018  &0.006  &0.001  &0.004  &-0.001  &-0.007  &-0.011  &-0.015  &0.005  &0.002  &-0.001  &-0.004  &-0.004  &-0.004  &-0.001  &-0.001  &0.004  &0.013 \\
4  &0.015  &0.028  &0.024  &0.018  &0.009  &0.001  &-0.005  &-0.012  &-0.017  &-0.020  &0.008  &0.002  &-0.005  &-0.010  &-0.012  &-0.012  &-0.003  &0.001  &0.015  &0.035 \\
5  &0.006  &0.023  &0.023  &0.020  &0.010  &0.001  &-0.003  &-0.011  &-0.019  &-0.023  &0.008  &-0.000  &-0.007  &-0.017  &-0.016  &-0.013  &-0.008  &0.005  &0.021  &0.046 \\
6  &-0.006  &0.006  &0.020  &0.021  &0.014  &0.007  &-0.001  &-0.010  &-0.018  &-0.024  &0.007  &0.001  &-0.009  &-0.015  &-0.019  &-0.013  &-0.008  &0.004  &0.020  &0.051 \\
7  &-0.015  &-0.008  &0.008  &0.016  &0.011  &0.007  &-0.002  &-0.001  &-0.009  &-0.017  &0.000  &-0.000  &-0.007  &-0.012  &-0.013  &-0.009  &-0.004  &0.004  &0.015  &0.031 \\
\hline
\end{tabularx}
 \end{adjustbox}
\end{sidewaystable*}

\begin{sidewaystable*}[htb]
\tiny
\caption{Statistical uncertainty correlation matrix of $(w, \cos\theta{\ell})$ versus $(\cos\theta_{v}, \chi)$ for the electron mode. The bins are defined in Section~\ref{differential-data}.}
\label{statcorr-table3-e}
\centering
\setlength{\tabcolsep}{4pt}
\renewcommand{\arraystretch}{1.5}
 \begin{adjustbox}{width=\textwidth}
\begin{tabularx}{0.8\linewidth}{r|rrrrrrrrrrrrrrrrrrrr}
\hline \hline
Bin &1      &2      &3      &4      &5      &6      &7      &8      &9      &10 &11      &12      &13      &14      &15      &16      &17      &18      &19      &20 \\
\hline
21  &0.022  &0.019  &0.017  &0.015  &0.006  &-0.006  &-0.015  &-0.023  &-0.024  &-0.026  &-0.003  &0.001  &0.004  &0.003  &-0.001  &-0.001  &0.000  &0.007  &0.004  &-0.001 \\
22  &0.009  &0.021  &0.027  &0.028  &0.023  &0.006  &-0.008  &-0.029  &-0.042  &-0.046  &-0.004  &0.004  &0.006  &0.001  &-0.002  &0.001  &0.004  &0.006  &0.007  &-0.006 \\
23  &0.002  &0.013  &0.018  &0.024  &0.023  &0.020  &0.008  &-0.010  &-0.038  &-0.054  &-0.004  &0.001  &0.008  &0.006  &0.003  &-0.003  &0.004  &0.006  &0.002  &-0.001 \\
24  &-0.001  &0.006  &0.006  &0.018  &0.020  &0.021  &0.016  &0.002  &-0.020  &-0.050  &-0.004  &0.005  &0.004  &0.005  &0.001  &0.003  &0.003  &0.009  &0.003  &-0.003 \\
25  &-0.003  &0.001  &0.001  &0.009  &0.010  &0.014  &0.011  &0.006  &-0.004  &-0.024  &-0.001  &0.000  &0.004  &0.002  &0.000  &0.003  &0.003  &0.004  &0.005  &-0.001 \\
26  &-0.004  &-0.005  &0.004  &0.001  &0.001  &0.007  &0.007  &0.009  &0.005  &-0.001  &0.000  &0.001  &0.002  &0.001  &0.002  &0.003  &0.004  &0.002  &0.000  &0.002 \\
27  &-0.003  &-0.003  &-0.001  &-0.005  &-0.003  &-0.001  &-0.002  &0.007  &0.014  &0.016  &0.005  &-0.001  &0.003  &-0.002  &-0.000  &0.003  &0.003  &0.000  &-0.001  &0.001 \\
\hline
\end{tabularx}
 \end{adjustbox}
\end{sidewaystable*}

\begin{sidewaystable*}[htb]
\tiny
\caption{Statistical correlation matrix of $(\cos\theta_{v}, \chi)$ versus $(\cos\theta_{v}, \chi)$ for the electron mode. The bins are defined in Section~\ref{differential-data}.}
\label{statcorr-table4-e}
\centering
\setlength{\tabcolsep}{4pt}
\renewcommand{\arraystretch}{1.5}
 \begin{adjustbox}{width=\textwidth}
\begin{tabularx}{0.8\linewidth}{r|rrrrrrrrrrrrrrrrrrrr}
\hline \hline
Bin &21      &22      &23      &24      &25      &26      &27      &28      &29      &30 &31      &32      &33      &34      &35      &36      &37      &38      &39      &40 \\
\hline
21  &1.000  &0.000  &0.000  &0.000  &0.000  &0.000  &0.000  &0.000  &0.000  &0.000  &-0.002  &-0.009  &-0.007  &0.005  &0.016  &0.018  &0.009  &-0.006  &-0.006  &-0.004 \\
22  &0.000  &1.000  &0.000  &0.000  &0.000  &0.000  &0.000  &0.000  &0.000  &0.000  &-0.014  &-0.009  &0.001  &0.008  &0.020  &0.019  &0.009  &-0.001  &-0.010  &-0.016 \\
23  &0.000  &0.000  &1.000  &0.000  &0.000  &0.000  &0.000  &0.000  &0.000  &0.000  &-0.019  &-0.010  &0.001  &0.010  &0.016  &0.017  &0.011  &0.004  &-0.009  &-0.018 \\
24  &0.000  &0.000  &0.000  &1.000  &0.000  &0.000  &0.000  &0.000  &0.000  &0.000  &-0.016  &-0.005  &0.005  &0.011  &0.006  &0.009  &0.006  &0.005  &-0.005  &-0.015 \\
25  &0.000  &0.000  &0.000  &0.000  &1.000  &0.000  &0.000  &0.000  &0.000  &0.000  &-0.009  &0.003  &0.004  &0.002  &-0.006  &-0.003  &0.003  &0.007  &-0.001  &-0.008 \\
26  &0.000  &0.000  &0.000  &0.000  &0.000  &1.000  &0.000  &0.000  &0.000  &0.000  &0.002  &0.005  &0.007  &-0.005  &-0.011  &-0.014  &-0.005  &0.003  &0.006  &0.001 \\
27  &0.000  &0.000  &0.000  &0.000  &0.000  &0.000  &1.000  &0.000  &0.000  &0.000  &0.011  &0.008  &0.003  &-0.009  &-0.019  &-0.019  &-0.009  &0.002  &0.009  &0.013 \\
\hline
\end{tabularx}
 \end{adjustbox}
\end{sidewaystable*}

\begin{sidewaystable*}[htb]
\tiny
\caption{Statistical uncertainty correlation matrix of $(w, \cos\theta_{\ell})$ versus $(w, \cos\theta_{\ell})$ for the muon mode. The bins are defined in Section~\ref{differential-data}.}
\label{statcorr-table1-mu}
\centering
\setlength{\tabcolsep}{4pt}
\renewcommand{\arraystretch}{1.5}
 \begin{adjustbox}{width=\textwidth}
\begin{tabularx}{0.8\linewidth}{r|rrrrrrrrrrrrrrrrrrrr}
\hline \hline
Bin &1      &2      &3      &4      &5      &6      &7      &8      &9      &10 &11      &12      &13      &14      &15      &16      &17      &18      &19      &20 \\
\hline
1  &1.000  &0.000  &0.000  &0.000  &0.000  &0.000  &0.000  &0.000  &0.000  &0.000  &0.020  &0.014  &0.005  &0.002  &-0.003  &-0.003  &-0.006  &-0.004  &-0.001  &0.006 \\
2  &0.000  &1.000  &0.000  &0.000  &0.000  &0.000  &0.000  &0.000  &0.000  &0.000  &0.034  &0.022  &0.009  &-0.005  &-0.007  &-0.009  &-0.009  &-0.003  &-0.000  &0.018 \\
3  &0.000  &0.000  &1.000  &0.000  &0.000  &0.000  &0.000  &0.000  &0.000  &0.000  &0.029  &0.019  &0.009  &-0.004  &-0.010  &-0.010  &-0.007  &-0.001  &0.007  &0.022 \\
4  &0.000  &0.000  &0.000  &1.000  &0.000  &0.000  &0.000  &0.000  &0.000  &0.000  &0.026  &0.016  &0.004  &-0.003  &-0.010  &-0.008  &-0.005  &-0.003  &0.009  &0.027 \\
5  &0.000  &0.000  &0.000  &0.000  &1.000  &0.000  &0.000  &0.000  &0.000  &0.000  &0.014  &0.012  &0.002  &-0.004  &-0.007  &-0.005  &-0.001  &0.001  &0.009  &0.023 \\
6  &0.000  &0.000  &0.000  &0.000  &0.000  &1.000  &0.000  &0.000  &0.000  &0.000  &-0.010  &0.009  &0.001  &-0.001  &-0.003  &-0.001  &0.000  &0.007  &0.005  &0.013 \\
7  &0.000  &0.000  &0.000  &0.000  &0.000  &0.000  &1.000  &0.000  &0.000  &0.000  &-0.020  &-0.004  &0.004  &-0.000  &0.001  &0.003  &0.003  &0.007  &0.007  &0.000 \\
\hline
\end{tabularx}
 \end{adjustbox}
\end{sidewaystable*}

\begin{sidewaystable*}[htb]
\tiny
\caption{Statistical uncertainty correlation matrix of $(\cos\theta_{v}, \chi)$ versus $(w,\cos\theta_{\ell})$ for the muon mode. The bins are defined in Section~\ref{differential-data}.}
\label{statcorr-table2-mu}
\centering
\setlength{\tabcolsep}{4pt}
\renewcommand{\arraystretch}{1.5}
 \begin{adjustbox}{width=\textwidth}
\begin{tabularx}{0.8\linewidth}{r|rrrrrrrrrrrrrrrrrrrr}
\hline \hline
Bin &21      &22      &23      &24      &25      &26      &27      &28      &29      &30 &31      &32      &33      &34      &35      &36      &37      &38      &39      &40 \\
\hline
1  &0.019  &0.012  &0.002  &-0.001  &-0.004  &-0.005  &0.000  &0.001  &0.003  &0.001  &-0.004  &0.005  &0.012  &0.014  &0.014  &0.013  &0.008  &-0.001  &-0.024  &-0.051 \\
2  &0.023  &0.019  &0.013  &0.004  &0.000  &-0.002  &-0.004  &-0.006  &-0.004  &-0.004  &0.003  &0.001  &0.005  &0.004  &0.006  &0.005  &0.004  &-0.000  &-0.007  &-0.019 \\
3  &0.015  &0.022  &0.018  &0.013  &0.003  &0.000  &-0.005  &-0.007  &-0.012  &-0.012  &0.004  &0.003  &-0.001  &-0.004  &-0.005  &-0.003  &-0.002  &0.003  &0.007  &0.013 \\
4  &0.012  &0.026  &0.023  &0.018  &0.009  &0.000  &-0.005  &-0.012  &-0.018  &-0.022  &0.005  &0.004  &-0.005  &-0.011  &-0.014  &-0.010  &-0.007  &0.002  &0.015  &0.041 \\
5  &0.009  &0.020  &0.021  &0.018  &0.010  &0.001  &-0.005  &-0.011  &-0.020  &-0.022  &0.007  &0.003  &-0.008  &-0.014  &-0.016  &-0.015  &-0.009  &0.006  &0.023  &0.043 \\
6  &-0.006  &0.011  &0.021  &0.017  &0.014  &0.006  &-0.003  &-0.011  &-0.017  &-0.022  &0.004  &0.000  &-0.007  &-0.014  &-0.018  &-0.013  &-0.009  &0.004  &0.023  &0.051 \\
7  &-0.017  &-0.009  &0.006  &0.012  &0.014  &0.008  &0.001  &-0.003  &-0.011  &-0.014  &0.000  &-0.001  &-0.007  &-0.009  &-0.011  &-0.009  &-0.005  &0.006  &0.019  &0.029 \\
\hline
\end{tabularx}
 \end{adjustbox}
\end{sidewaystable*}

\begin{sidewaystable*}[htb]
\tiny
\caption{Statistical uncertainty correlation matrix of $(w, \cos\theta{\ell})$ versus $(\cos\theta_{v}, \chi)$ for the muon mode. The bins are defined in Section~\ref{differential-data}.}
\label{statcorr-table3-mu}
\centering
\setlength{\tabcolsep}{4pt}
\renewcommand{\arraystretch}{1.5}
 \begin{adjustbox}{width=\textwidth}
\begin{tabularx}{0.8\linewidth}{r|rrrrrrrrrrrrrrrrrrrr}
\hline \hline
Bin &1      &2      &3      &4      &5      &6      &7      &8      &9      &10 &11      &12      &13      &14      &15      &16      &17      &18      &19      &20 \\
\hline
21  &0.019  &0.023  &0.015  &0.012  &0.009  &-0.006  &-0.017  &-0.020  &-0.023  &-0.024  &-0.002  &0.002  &0.006  &0.002  &-0.002  &-0.002  &0.005  &0.005  &0.002  &-0.003 \\
22  &0.012  &0.019  &0.022  &0.026  &0.020  &0.011  &-0.009  &-0.024  &-0.039  &-0.043  &-0.004  &0.001  &0.004  &0.005  &0.001  &0.000  &0.006  &0.004  &0.006  &-0.003 \\
23  &0.002  &0.013  &0.018  &0.023  &0.021  &0.021  &0.006  &-0.013  &-0.035  &-0.050  &-0.003  &0.001  &0.006  &0.004  &0.006  &-0.000  &0.007  &0.006  &0.002  &-0.005 \\
24  &-0.001  &0.004  &0.013  &0.018  &0.018  &0.017  &0.012  &0.001  &-0.018  &-0.042  &-0.002  &0.002  &0.006  &0.009  &0.003  &0.001  &0.005  &0.007  &-0.000  &-0.004 \\
25  &-0.004  &0.000  &0.003  &0.009  &0.010  &0.014  &0.014  &0.007  &-0.004  &-0.020  &-0.001  &0.002  &0.006  &0.004  &0.001  &0.002  &0.003  &0.004  &0.003  &-0.000 \\
26  &-0.005  &-0.002  &0.000  &0.000  &0.001  &0.006  &0.008  &0.008  &0.007  &-0.003  &0.004  &0.004  &0.001  &0.001  &0.002  &0.002  &0.002  &0.002  &-0.001  &0.001 \\
27  &0.000  &-0.004  &-0.005  &-0.005  &-0.005  &-0.003  &0.001  &0.007  &0.014  &0.017  &0.003  &0.001  &0.002  &0.001  &-0.001  &-0.002  &0.000  &0.001  &0.002  &0.003 \\
\hline
\end{tabularx}
 \end{adjustbox}
\end{sidewaystable*}

\begin{sidewaystable*}[htb]
\tiny
\caption{Statistical correlation matrix of $(\cos\theta_{v}, \chi)$ versus $(\cos\theta_{v}, \chi)$ for the muon mode. The bins are defined in Section~\ref{differential-data}.}
\label{statcorr-table4-mu}
\centering
\setlength{\tabcolsep}{4pt}
\renewcommand{\arraystretch}{1.5}
 \begin{adjustbox}{width=\textwidth}
\begin{tabularx}{0.8\linewidth}{r|rrrrrrrrrrrrrrrrrrrr}
\hline \hline
Bin &21      &22      &23      &24      &25      &26      &27      &28      &29      &30 &31      &32      &33      &34      &35      &36      &37      &38      &39      &40 \\
\hline
21  &1.000  &0.000  &0.000  &0.000  &0.000  &0.000  &0.000  &0.000  &0.000  &0.000  &-0.006  &-0.007  &-0.004  &0.005  &0.017  &0.017  &0.003  &-0.005  &-0.005  &-0.001 \\
22  &0.000  &1.000  &0.000  &0.000  &0.000  &0.000  &0.000  &0.000  &0.000  &0.000  &-0.014  &-0.011  &-0.000  &0.010  &0.018  &0.021  &0.012  &0.001  &-0.009  &-0.016 \\
23  &0.000  &0.000  &1.000  &0.000  &0.000  &0.000  &0.000  &0.000  &0.000  &0.000  &-0.018  &-0.009  &0.004  &0.009  &0.017  &0.018  &0.011  &0.003  &-0.011  &-0.018 \\
24  &0.000  &0.000  &0.000  &1.000  &0.000  &0.000  &0.000  &0.000  &0.000  &0.000  &-0.016  &-0.006  &0.006  &0.009  &0.004  &0.005  &0.009  &0.004  &-0.004  &-0.016 \\
25  &0.000  &0.000  &0.000  &0.000  &1.000  &0.000  &0.000  &0.000  &0.000  &0.000  &-0.010  &-0.002  &0.005  &0.005  &-0.004  &-0.004  &0.001  &0.007  &-0.001  &-0.007 \\
26  &0.000  &0.000  &0.000  &0.000  &0.000  &1.000  &0.000  &0.000  &0.000  &0.000  &0.005  &0.007  &0.007  &-0.003  &-0.013  &-0.015  &-0.003  &0.003  &0.005  &-0.000 \\
27  &0.000  &0.000  &0.000  &0.000  &0.000  &0.000  &1.000  &0.000  &0.000  &0.000  &0.011  &0.010  &0.001  &-0.008  &-0.019  &-0.019  &-0.007  &0.004  &0.010  &0.011 \\
\hline
\end{tabularx}
 \end{adjustbox}
\end{sidewaystable*}

\begin{sidewaystable*}[htb]
\tiny
\caption{Systematic uncertainty correlation matrix of $(w, \cos\theta_{\ell})$ versus $(w, \cos\theta_{\ell})$. The bins are defined in Section~\ref{differential-data}.}
\label{syscorr-table1}
\centering
\setlength{\tabcolsep}{5pt}
\renewcommand{\arraystretch}{1.5}
 \begin{adjustbox}{width=\textwidth}
\begin{tabularx}{0.8\linewidth}{r|rrrrrrrrrrrrrrrrrrrr}
\hline \hline
Bin &1      &2      &3      &4      &5      &6      &7      &8      &9      &10 &11      &12      &13      &14      &15      &16      &17      &18      &19      &20 \\
\hline
1  &1.000  &0.872  &0.755  &0.682  &0.656  &0.612  &0.555  &0.500  &0.487  &0.409  &0.786  &0.699  &0.633  &0.621  &0.623  &0.628  &0.630  &0.640  &0.653  &0.662 \\
2  &0.872  &1.000  &0.955  &0.899  &0.878  &0.850  &0.819  &0.779  &0.763  &0.680  &0.934  &0.891  &0.855  &0.857  &0.865  &0.872  &0.872  &0.879  &0.889  &0.896 \\
3  &0.755  &0.955  &1.000  &0.978  &0.957  &0.936  &0.917  &0.891  &0.885  &0.819  &0.952  &0.946  &0.933  &0.943  &0.951  &0.957  &0.957  &0.963  &0.969  &0.974 \\
4  &0.682  &0.899  &0.978  &1.000  &0.985  &0.967  &0.952  &0.943  &0.931  &0.854  &0.934  &0.950  &0.952  &0.959  &0.968  &0.977  &0.979  &0.984  &0.986  &0.990 \\
5  &0.656  &0.878  &0.957  &0.985  &1.000  &0.992  &0.978  &0.950  &0.941  &0.875  &0.925  &0.957  &0.968  &0.980  &0.986  &0.987  &0.989  &0.990  &0.993  &0.992 \\
6  &0.612  &0.850  &0.936  &0.967  &0.992  &1.000  &0.994  &0.963  &0.948  &0.879  &0.906  &0.950  &0.970  &0.982  &0.987  &0.986  &0.988  &0.987  &0.989  &0.985 \\
7  &0.555  &0.819  &0.917  &0.952  &0.978  &0.994  &1.000  &0.979  &0.967  &0.902  &0.884  &0.938  &0.967  &0.982  &0.987  &0.986  &0.986  &0.982  &0.983  &0.976 \\
\hline
\end{tabularx}
 \end{adjustbox}
\end{sidewaystable*}

\begin{sidewaystable*}[htb]
\tiny
\caption{Systematic correlation matrix $(\cos\theta_{v}, \chi)$ and $(w,\cos\theta_{\ell})$ distribution. The bins are defined in Section~\ref{differential-data}.}
\label{syscorr-table2}
\centering
\setlength{\tabcolsep}{5pt}
\renewcommand{\arraystretch}{1.5}
 \begin{adjustbox}{width=\textwidth}
\begin{tabularx}{0.8\linewidth}{r|rrrrrrrrrrrrrrrrrrrr}
\hline \hline
Bin &21      &22      &23      &24      &25      &26      &27      &28      &29      &30 &31      &32      &33      &34      &35      &36      &37      &38      &39      &40 \\
\hline
1  &0.508  &0.763  &0.877  &0.919  &0.954  &0.970  &0.983  &0.975  &0.981  &0.951  &0.851  &0.920  &0.958  &0.980  &0.981  &0.974  &0.970  &0.964  &0.961  &0.950 \\
2  &0.524  &0.783  &0.895  &0.938  &0.967  &0.981  &0.991  &0.982  &0.983  &0.942  &0.866  &0.930  &0.964  &0.984  &0.987  &0.983  &0.980  &0.976  &0.973  &0.964 \\
3  &0.558  &0.814  &0.921  &0.963  &0.983  &0.990  &0.994  &0.986  &0.981  &0.926  &0.886  &0.941  &0.968  &0.985  &0.991  &0.991  &0.991  &0.988  &0.987  &0.981 \\
4  &0.591  &0.842  &0.942  &0.982  &0.991  &0.991  &0.989  &0.980  &0.969  &0.901  &0.902  &0.947  &0.968  &0.980  &0.987  &0.991  &0.993  &0.992  &0.992  &0.990 \\
5  &0.638  &0.880  &0.967  &0.992  &0.990  &0.981  &0.973  &0.965  &0.958  &0.896  &0.918  &0.950  &0.961  &0.977  &0.985  &0.992  &0.993  &0.995  &0.995  &0.996 \\
6  &0.686  &0.911  &0.984  &0.996  &0.988  &0.973  &0.960  &0.945  &0.938  &0.873  &0.934  &0.953  &0.957  &0.971  &0.979  &0.986  &0.988  &0.991  &0.993  &0.996 \\
7  &0.731  &0.938  &0.993  &0.991  &0.982  &0.964  &0.946  &0.922  &0.916  &0.848  &0.944  &0.953  &0.950  &0.963  &0.972  &0.978  &0.980  &0.984  &0.987  &0.991 \\
\hline
\end{tabularx}
 \end{adjustbox}
\end{sidewaystable*}

\begin{sidewaystable*}[htb]
\tiny
\caption{Systematic correlation matrix $(w, \cos\theta{\ell})$ and $(\cos\theta_{v}, \chi)$ distributions. The label for each bin is defined in Section~\ref{differential-data}.}
\label{syscorr-table3}
\centering
\setlength{\tabcolsep}{5pt}
\renewcommand{\arraystretch}{1.5}
 \begin{adjustbox}{width=\textwidth}
\begin{tabularx}{0.8\linewidth}{r|rrrrrrrrrrrrrrrrrrrr}
\hline \hline
Bin &1      &2      &3      &4      &5      &6      &7      &8      &9      &10 &11      &12      &13      &14      &15      &16      &17      &18      &19      &20 \\
\hline
21  &0.508  &0.524  &0.558  &0.591  &0.638  &0.686  &0.731  &0.756  &0.781  &0.840  &0.666  &0.668  &0.651  &0.633  &0.608  &0.609  &0.634  &0.657  &0.670  &0.665 \\
22  &0.763  &0.783  &0.814  &0.842  &0.880  &0.911  &0.938  &0.952  &0.965  &0.993  &0.900  &0.900  &0.886  &0.869  &0.849  &0.851  &0.871  &0.890  &0.902  &0.899 \\
23  &0.877  &0.895  &0.921  &0.942  &0.967  &0.984  &0.993  &0.997  &0.996  &0.976  &0.974  &0.975  &0.965  &0.955  &0.942  &0.944  &0.957  &0.969  &0.975  &0.974 \\
24  &0.919  &0.938  &0.963  &0.982  &0.992  &0.996  &0.991  &0.979  &0.968  &0.925  &0.989  &0.989  &0.981  &0.975  &0.968  &0.970  &0.978  &0.985  &0.985  &0.989 \\
25  &0.954  &0.967  &0.983  &0.991  &0.990  &0.988  &0.982  &0.968  &0.956  &0.912  &0.991  &0.992  &0.993  &0.990  &0.987  &0.987  &0.992  &0.990  &0.990  &0.990 \\
26  &0.970  &0.981  &0.990  &0.991  &0.981  &0.973  &0.964  &0.947  &0.936  &0.887  &0.984  &0.985  &0.991  &0.990  &0.990  &0.991  &0.992  &0.986  &0.984  &0.984 \\
27  &0.983  &0.991  &0.994  &0.989  &0.973  &0.960  &0.946  &0.928  &0.915  &0.861  &0.977  &0.978  &0.986  &0.988  &0.991  &0.992  &0.989  &0.981  &0.977  &0.977 \\
\hline
\end{tabularx}
 \end{adjustbox}
\end{sidewaystable*}

\begin{sidewaystable*}[htb]
\tiny
\caption{Systematic uncertainty correlation matrix between $(\cos\theta_{v}, \chi)$ and $(\cos\theta_{v}, \chi)$ distribution. The bins are defined in Section~\ref{differential-data}.}
\label{syscorr-table4}
\centering
\setlength{\tabcolsep}{5pt}
\renewcommand{\arraystretch}{1.5}
 \begin{adjustbox}{width=\textwidth}
\begin{tabularx}{0.8\linewidth}{r|rrrrrrrrrrrrrrrrrrrr}
\hline \hline
Bin &21      &22      &23      &24      &25      &26      &27      &28      &29      &30 &31      &32      &33      &34      &35      &36      &37      &38      &39      &40 \\
\hline
21  &1.000  &0.998  &0.987  &0.968  &0.949  &0.929  &0.911  &0.894  &0.873  &0.813  &0.953  &0.957  &0.969  &0.978  &0.985  &0.984  &0.975  &0.964  &0.959  &0.954 \\
22  &0.998  &1.000  &0.995  &0.981  &0.964  &0.946  &0.928  &0.910  &0.890  &0.830  &0.966  &0.969  &0.979  &0.986  &0.991  &0.991  &0.985  &0.975  &0.970  &0.966 \\
23  &0.987  &0.995  &1.000  &0.995  &0.983  &0.968  &0.952  &0.933  &0.914  &0.855  &0.981  &0.983  &0.989  &0.993  &0.995  &0.996  &0.992  &0.987  &0.982  &0.981 \\
24  &0.968  &0.981  &0.995  &1.000  &0.994  &0.984  &0.970  &0.950  &0.934  &0.878  &0.990  &0.990  &0.991  &0.991  &0.991  &0.992  &0.992  &0.991  &0.988  &0.990 \\
25  &0.949  &0.964  &0.983  &0.994  &1.000  &0.996  &0.987  &0.974  &0.959  &0.912  &0.995  &0.995  &0.992  &0.990  &0.986  &0.987  &0.990  &0.995  &0.994  &0.995 \\
26  &0.929  &0.946  &0.968  &0.984  &0.996  &1.000  &0.997  &0.988  &0.978  &0.939  &0.995  &0.995  &0.989  &0.984  &0.977  &0.978  &0.986  &0.993  &0.994  &0.995 \\
27  &0.911  &0.928  &0.952  &0.970  &0.987  &0.997  &1.000  &0.996  &0.991  &0.962  &0.991  &0.991  &0.983  &0.976  &0.967  &0.968  &0.978  &0.986  &0.990  &0.991 \\
\hline
\end{tabularx}
 \end{adjustbox}
\end{sidewaystable*}

\end{appendix}

\begin{thebibliography}{99}
\bibitem{cite-chargeconjugation}
Throughout this note charge-conjugate decay modes are
implied.
\end{thebibliography}
\end{document}